\documentclass[%
 reprint,
superscriptaddress,
 amsmath,amssymb,
 aps,
]{revtex4-1}

\usepackage{graphicx}% Include figure files
\usepackage{dcolumn}% Align table columns on decimal point
\usepackage{xcolor}
\usepackage{soul}
\usepackage{here}
\usepackage{dcolumn}% Align table columns on decimal point
\usepackage{bm}% bold math
\usepackage{lipsum}  
\begin{document}

%\preprint{ApS/123-QED}

%\title{\textcolor{blue}{Trap model of Maxwell glass
%rheology in protein condensate}}
\title{Theory of rheology and aging of protein condensates}% Force line breaks with \\
%\thanks{A footnote to the article title}%

\author{Ryota Takaki}
\affiliation{Max Planck Institute for the Physics of Complex Systems, Nöthnitzer Str.38, 01187 Dresden, Germany}
%\altaffiliation[Also at ]{physics Department, University Texas at Austin.}%Lines break automatically or can be forced with \\
\author{Louise Jawerth}%
\affiliation{Soft Matter Physics, Huygens-Kamerlingh Onnes Laboratory, Leiden University, P.O. Box 9504, 2300 RA Leiden, The Netherlands}

\author{Marko Popovi\'c}
\email{Corresponding author: mpopovic@pks.mpg.de}
\affiliation{Max Planck Institute for the Physics of Complex Systems, Nöthnitzer Str.38, 01187 Dresden, Germany}

\author{Frank Jülicher}
\email{Corresponding author: julicher@pks.mpg.de}
\affiliation{Max Planck Institute for the Physics of Complex Systems, Nöthnitzer Str.38, 01187 Dresden, Germany}

% \affil[1]{Max Planck Institute for the Physics of Complex Systems, Nöthnitzer Str.38, 01187 Dresden, Germany}
% \affil[2]{Soft Matter Physics, Huygens-Kamerlingh Onnes Laboratory, Leiden University,
% P.O. Box 9504, 2300 RA Leiden, The Netherlands}

%\collaboration{MUSO Collaboration}%\noaffiliation

% \author{}
%  \homepage{http://www.Second.institution.edu/~Charlie.Author}
% \affiliation{ \\
%  Second institution and/or address\\
%  This line break forced% with \\
% }%
% \affiliation{
%  Third institution, the second for Charlie Author
% }%
% \author{Delta Author}
% \affiliation{%
%  Authors' institution and/or address\\
%  This line break forced with \textbackslash\textbackslash
% }%

%\collaboration{CLEO Collaboration}%\noaffiliation

\date{\today}% It is always \today, today,
             %  but any date may be explicitly specified

\begin{abstract}

Biological condensates are assemblies of proteins and nucleic acids that form membraneless compartments in cells and play essential roles in cellular functions. In many cases they exhibit the physical properties of liquid droplets that coexist in a surrounding fluid. Recently, quantitative studies on the material properties of biological condensates have become available, revealing complex material properties~\cite{jawerth2020protein,alshareedah2021programmable}. In vitro experiments have shown that protein condensates exhibit time dependent material properties, similar to aging in glasses. To understand this phenomenon from a theoretical perspective, we develop a rheological model based on the physical picture of protein diffusion and stochastic binding inside condensates. The complex nature of protein interactions is captured by a distribution of binding energies, incorporated in a trap model originally developed to study glass transitions~\cite{bouchaud1992weak}. Our model can describe diffusion of constituent particles, as well as the material response to time-dependent forces, and it recapitulates the age dependent relaxation time of Maxwell glass observed experimentally both in active and passive rheology. We derive fluctuation-response relations of our model in which the relaxation function does not obey time translation invariance. Our study sheds light on the complex material properties of biological condensates and provides a theoretical framework for understanding their aging behavior.
\end{abstract}

\pacs{Valid pACS appear here}% pACS, the physics and Astronomy
                             % Classification Scheme.
%\keywords{Suggested keywords}%Use showkeys class option if keyword
                              %display desired
\maketitle

%\tableofcontents

\section{\label{sec:level1}Introduction}
The formation of biological condensates by phase separation of proteins and nucleic acids in the cell   has became a new paradigm in molecular biology over the last decade~\cite{brangwynne2009germline,hyman2014liquid,banani2017biomolecular}. Such condensates provide membrane-less biochemical compartments with liquid-like properties. They typically exhibit a spherical shape  to minimize the surface tension and have properties of droplets in a fluid environment. Recent studies suggest that rheological properties of biomolecular condensates can be considerably richer than those of simple liquids~\cite{jawerth2020protein,ghosh2021shear,riback2022viscoelastic}, which may have biological consequences~\cite{riback2022viscoelastic,patel2015liquid,shin2017liquid,franzmann2018phase}. 

Recently, the rheological property of RNA associated condensates of PGL-3 and FUS protein condensates were studied in vitro using active and passive microrheology~\cite{jawerth2020protein}. The study revealed  time-dependent material properties of these protein condensates, summarized as follows: (1) The rheological properties of the condensates depend on the waiting time ($t_w$) between droplet formation and experiment; they are well fit by a Maxwell fluid model with elastic behavior on short time scales up to the relaxation time ($\tau_c$) and liquid behavior at the longer time scales. (2) The relaxation time, $\tau_c$, of the Maxwell fluid increases for longer waiting time $t_w$. The increase of $\tau_c$ is associated with an strong increase of viscosity, while the change of elasticity is small. (3) Various quantities reflecting the material property, such as complex modulus and mean squared displacement, collapse on a master curve upon rescaling of frequency and modulus for different $t_w$.  These time-dependent rheological properties suggest that the rheology of the protein condensates is an aging Maxwell fluid, termed Maxwell glass, referring to aging phenomena in glassy materials~\cite{berthier2011theoretical,kirkpatrick2015colloquium}. 

Viscoelastic properties of condensates have been reported in multiple experimental studies. Alshareedah {\it et al.}~\cite{alshareedah2021programmable} found that condensate viscoelasticity can be modulated by varying aminoacid sequence of condensate-forming proteins. Ghosh {\it et al.}~\cite{ghosh2021shear} investigated the relationship between condensate rheology and fusion dynamics showing that shorter relaxation times lead to faster fusion.  %Since the determination of bulk material properties is affected by condensate surface tension, Refs.\cite{jawerth2018salt} and~\cite{zhou2020determination} developed an approach to determine bulk material properties from rheology data obtained for condensates. 
Theory on viscoelastic condensates has addressed the shape dynamics of condensate droplets \cite{zhou2021shape}, as well as salt dependence of viscoelastic material properties \cite{zhou2021viscoelasticity}. 
A two fluid model describing the transition from a liquid to an elastic droplet was proposed to discuss the observed solid-like condensate behaviours~\cite{meng2023indissoluble}. %is studied theoretically.
Shen {\it et al.}~\cite{Shen2022.08.15.503964} reported the spatially heterogeneous condensate organisation during the transition from a liquid to a solid state in an aging condensate.

Aging and complex rheology of non-biological materials has a long history of research~\cite{chen2010rheology} due to its abundance and close connection to daily life~\cite{doraiswamy2002origins}. A comprehensive experimental study of aging materials by Struik dates back to the 1970s~\cite{struik1977physical}. More recently, aging colloidal glasses have been studied using microrheology~\cite{jabbari2007fluctuation}. The soft glassy rheology (SGR) model has been developed to describe the aging and rheology of soft materials~\cite{sollich1997rheology,sollich1998rheological,fielding2000aging}, based on seminal works by Bouchaud and coworkers~\cite{bouchaud1992weak,monthus1996models}. However, in the aging regime, the SGR model exhibits a solid-like behavior which does not describe an aging Maxwell fluid. %Based on the trap model, Sollich {\it et al.} developed the soft glassy rheology model to describe the rheology of soft glassy materials and their aging properties~\cite{fielding2000aging}. 
Recently, Lin~\cite{lin2022modeling} proposed a related mean-field model for condensate aging, based on the assumption of strongly correlated transitions between trap energies, in contrast to the soft glassy rheology model. Calculating the linear response function in this model yields a linear aging of condensate relaxation time-scale.

In this work, we develop a mean-field model of aging biological condensates that can describe their time-dependent material properties, observed in experiments.  We clarify how the aging of the protein condensates is reflected in active and passive microrheology. Active and passive rheology methods are illustrated in Fig.\ref{Fig:schematics}a. The structure of the paper is as follows. In section~\ref{sec:Model}, we propose a mean-field model to describe the binding and unbinding of diffusive elements inside the protein condensates. 
%We note that this diffusion of microscopic elements differs from the diffusion of tracer beads in passive rheology discussed in section~\ref{sec:PR}. 
Using the unbound probability of elements in condenates, we write the constitutive equation of the aging Maxwell fluid, leading to the relaxation function for Maxwell glass (section \ref{sec:CEMGa}). In section~\ref{sec:AC}, we examine the time-dependent rheology of the model using active rheology and propose the time-dependent complex modulus. Finally, in section~\ref{sec:PR}, we derive fluctuation-response relations between response functions and mean squared displacement of the diffusive elements, which can be employed in passive rheology experiments. We conclude with a discussion of our results. For readers unfamiliar with the subject, we have included an introduction to the rheology of aging materials in Appendix \ref{sec:rheology}, which summarizes the essential concepts employed throughout the paper.

%\section{Active Rheology}
\section{\label{sec:Model} Trap model of condensate aging}
\begin{figure}
\begin{center}
%\scalebox{1.5}[1.5]{
\includegraphics[width=0.5\textwidth]{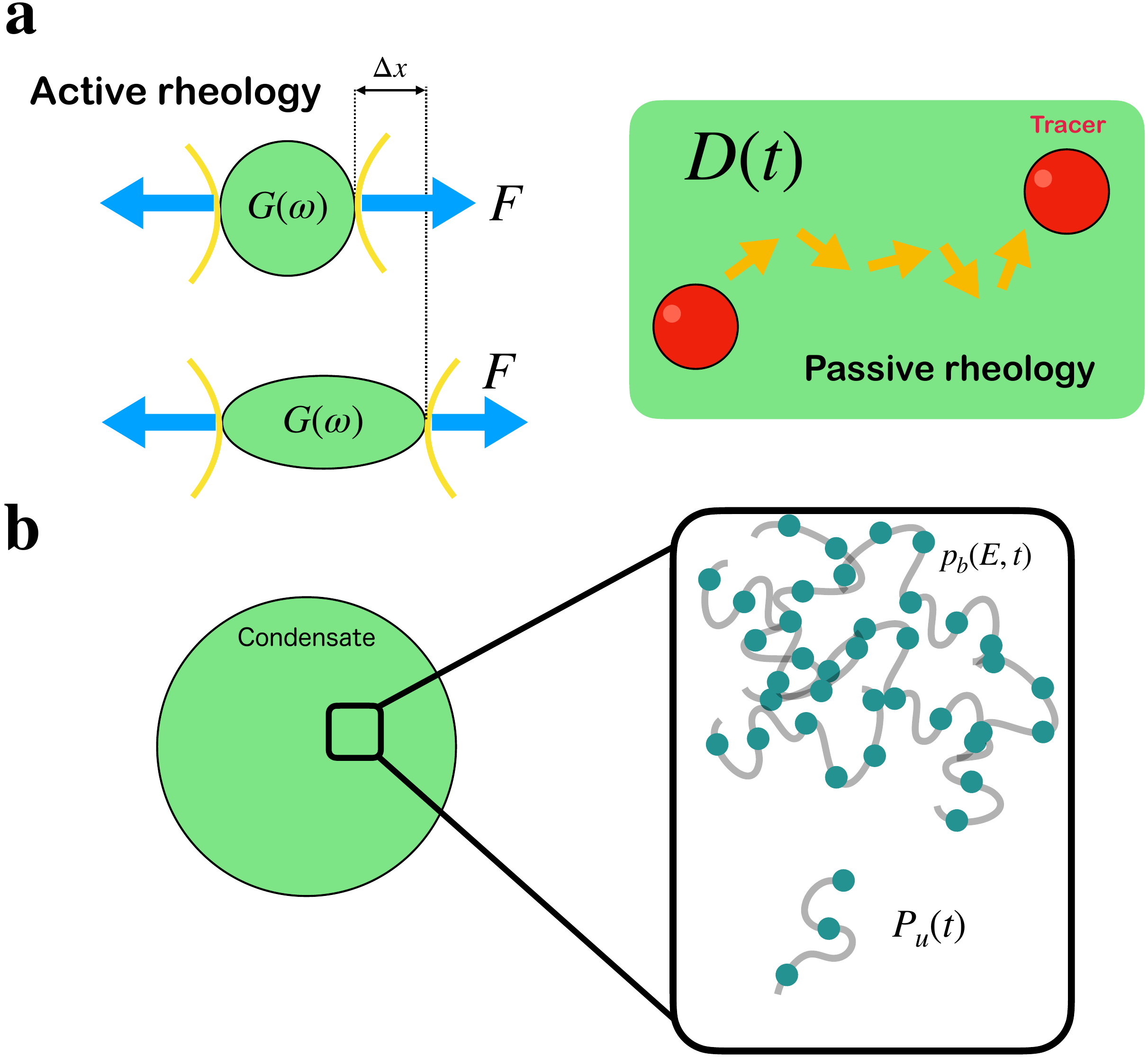}
%}
\end{center}
\caption{\label{Fig:schematics}  Schematics of the model and methods of microrheology. {\bf (a)} {\bf Left}: Schematics of active rheology. The external force ($F$) is applied to the protein condensates (green spheres) having complex modulus $G(\omega)$ using optical tweezers (yellow). The relation between strain and stress gives the material property of condensates. {\bf Right}: Schematics of passive rheology. The motion of the tracer element (red) embedded into the condensate is tracked. The element's mean square displacement encodes the condensate's material property, which manifests as diffusion coefficient $D(t)$. {\bf (b)} Schematics of the model. The diffusing element takes two states. One is the bound state, where chemical cross-links are densely connected at the reaction sites (green circles) so that the diffusion of the elements is hindered. The other is the unbound state, where the the diffusing elements can freely undergo diffusive motion. We denote the probability density of the bound state as $p_b(E,t)$ and the probability of the unbound state as $P_u(t)$ (see the main text for the detail).   }
\end{figure}

We introduce a mean-field model of an aging protein condensate composed of cross-linked elements forming an elastic network. These elements occasionally unbind and freely diffuse before attaching at a new location, see Fig.\ref{Fig:schematics}b. Dynamics of unbinding is determined by the binding energy $E$ of individual cross-links. To describe cross-linking of large proteins in a complex environment we draw binding energies from a distribution $\rho(E)$.  
The state of each cross-linker at time $t$ is described by probabilities $p_b(E, t)$ and $P_u(t)$ to find it bound with energy $E$ or unbound, respectively. In our mean-field model individual cross-linker probabilities also represent the fraction of all cross-linkers in the corresponding state.  The dynamical equations for these probabilities are

\begin{subequations}
\label{eq:ME2}
  \begin{align} 
  \label{eq:ME2a}
   &\frac{1}{\Gamma_0}\frac{\partial  p_b(E,t)}{\partial t}=-p_b(E,t) e^{-\beta E}+P_u(t)\rho(E) ,\\
   \label{eq:ME2b}
   &\frac{1}{\Gamma_0}\frac{\partial P_u(t)}{\partial t}  =- P_u(t) + \int_0^{\infty}dE p_b(E,t)e^{-\beta E},
  \end{align}
\end{subequations}
where $\beta \equiv 1/k_B T$, with temperature $T$ and Boltzmann constant $k_B$. $T$ is the temperature of the heat bath to which the condensates are coupled.  
%$\beta$ is the inverse temperature and $\beta_0$ is a parameter to determine the energy landscape of the binding of diffusing elements. 

Eq.(\ref{eq:ME2}) is an extension of trap model by Bouchaud~\cite{bouchaud1992weak,monthus1996models}. %The physical meaning of the Eq.(\ref{eq:ME2}) is explained as follows. 
The first term of the right-hand side in Eq.(\ref{eq:ME2a}) describes the transition from a bound state with energy $-E$ to the unbound state at $E=0$, which occur at a rate $\Gamma_0 e^{-\beta E}$, where $\Gamma_0$ is a rate parameter and binding energy $E>0$ is positive. The second term describes transitions from the unbound state to a bound state which occur at a density $\rho(E)$. 
%In  Eq.(\ref{eq:ME2b}), the first term describes transitions from the unbound state to a bound state, and the last term is the total probability flux from the bound state having energy $E$ to the unbound state, which is integrated over all the possible energies in the bound state. 

\begin{figure}
\begin{center}
%\scalebox{1.5}[1.5]{
\includegraphics[width=0.45\textwidth]{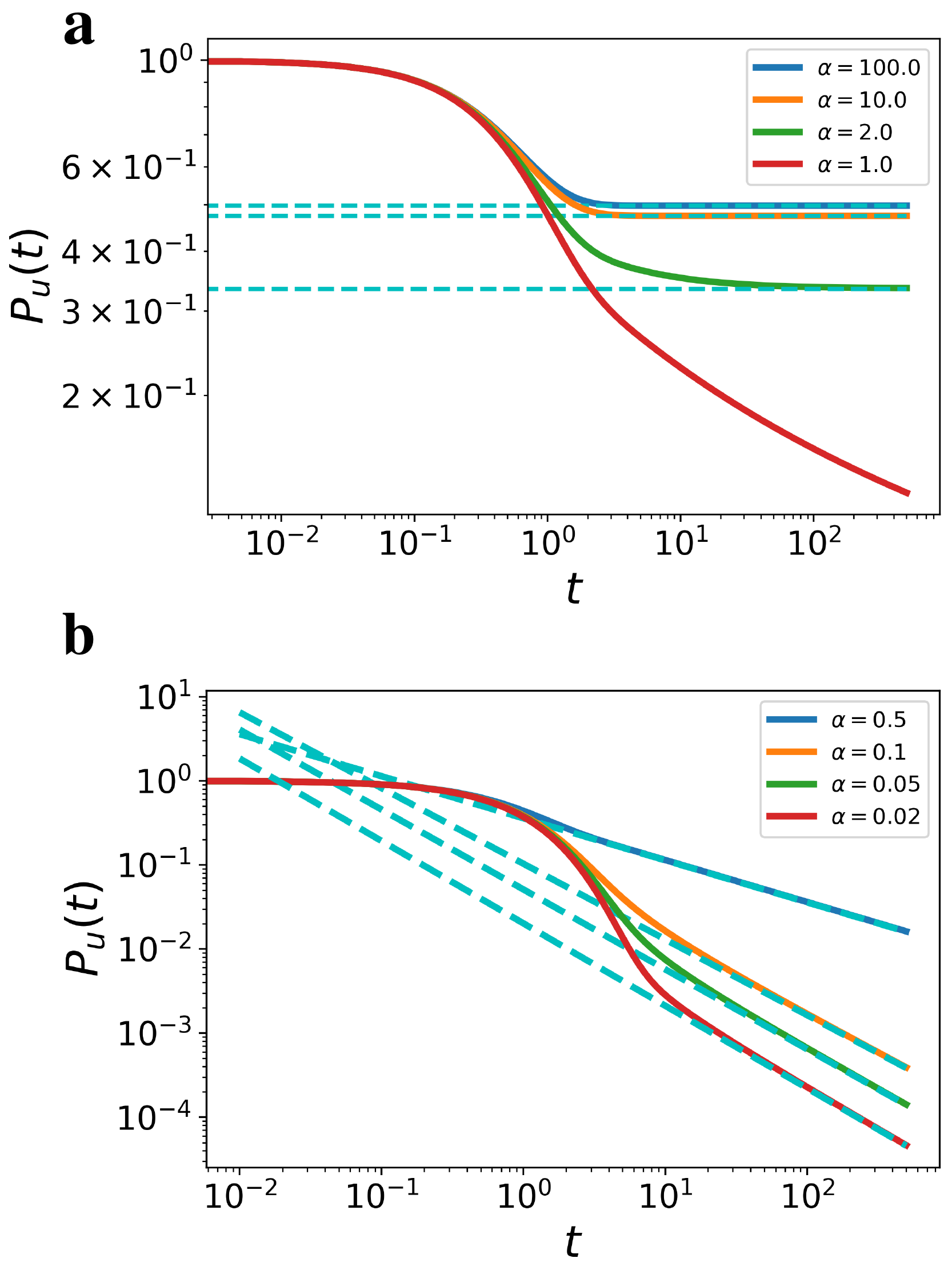}
%}
\end{center}
\caption{\label{Fig:Puplots} Dynamics of the unbound probability $P_u(t)$. Solid lines are numerically obtained from Eq.(\ref{eq:ME2}) and dashed lines are analytical solutions from Eq.(\ref{eq:Pusteady}) or Eq.(\ref{eq:pulong1}). The initial condition $p_b(E,0)=0\ (P_u(t=0)=1)$. We set $\Gamma_0=1$, which characterizes the time scale of the initial relaxation ($t \approx 1/\Gamma_0$), and measure the time ($t$) in the unit of $1/\Gamma_0$. We fix $\beta_0$ to $1$ and vary $\beta$. {\bf (a)} $P_u(t)$ for $\alpha \geq 1$. The dashed lines in cyan are the analytical solutions from Eq.(\ref{eq:Pusteady}). The equilibrium solutions exist for $\alpha > 1$. {\bf (b)} $P_u(t)$ for $\alpha < 1$.  $P_u(t)$ shows aging dynamics (slow relaxation) for long time regime. The dashed lines in cyan are the analytical solutions from Eq.(\ref{eq:pulong1}).}
\end{figure}

Here, we choose an exponential distribution of binding energies, $\rho(E)=\beta_0 e^{-\beta_0 E}$, which can describe both equilibrium and aging regimes of the model \cite{bouchaud1992weak}.
The parameter, $\alpha \equiv \beta_0/\beta$ controls qualitatively different solutions of Eq.(\ref{eq:ME2}). %\textcolor{blue}{The regimes $\alpha > 1$ and $\alpha < 1$ may be qualitatively associated with weak and strong binding regimes comparing to the unbinding for cross-linking proteins due to the thermal fluctuation.}  
For $\alpha > 1$,
the rate at which  bound states are populated decays faster with $E$ than the unbinding rate, and the system relaxes to an equilibrium  steady state with $p_b^{\rm eq}(E)\sim \rho(E)\exp(\beta E)$, 
and 
\begin{equation} 
\begin{split}
\label{eq:Pusteady}
P_u^{\rm eq}=\frac{\alpha-1}{2 \alpha -1 } \quad ,
\end{split}
\end{equation}
see Appendix~\ref{sec:agingsol}. As shown in \cite{bouchaud1992weak}, for $0 <\alpha < 1$ the rate at which bound state are populated decays slower with $E$ than the unbinding rate, so that $p_b^\text{eq}(E)$ is no longer normalizable and the equilibrium state of Eq.(\ref{eq:ME2}) does not exist.
The probability $P_u(t)$ vanishes asymptotically as%We discuss the relaxation dynamics to  equilibrium in Appendix~\ref{sec:Relaxation}, which shows that $\alpha$ determines the relaxation dynamics of $P_u(t)$. 
%We solve Eq.(\ref{eq:ME2}) formally in Laplace space (Appendix~\ref{sec:agingsol}). Going back to time domain, we obtain the long time behavior of $P_u(t)$.  For $0<\alpha<1$, when the system does not relax to equilibrium we find for large $t$ 
\begin{equation} 
\begin{split}
\label{eq:pulong1}
P_u(t) \simeq \kappa \big(\Gamma_0 t\big)^{\alpha-1} ; \ \ \ \kappa=\frac{1}{\alpha}\frac{\sin \big(\alpha \pi \big) }{\pi \Gamma [\alpha ] },
\end{split}
\end{equation}
as derived in Appendix~\ref{sec:agingsol}. Here $ \Gamma [\alpha]$ denotes the Gamma function. 
% The coefficient of the power can be expanded for $\alpha \simeq 0$ as $\alpha + o(\alpha)$, thus, 
% \begin{equation} 
% \begin{split}
% \label{eq:Puaging}
% P_u(t) \simeq \alpha (\Gamma_0t)^{\alpha-1},
% \end{split}
% \end{equation}
% where $\alpha \ll 1$.
%Eq.(\ref{eq:pulong1}) gives simple description in the aging regime for $P_u(t)$. 
%revealing that $P_u(t)$ decreases as power law with exponent $\alpha-1$. 
Fig.\ref{Fig:Puplots} shows $P_u(t)$ for initial condition $P_u(t=0)= 1$ evaluated for different values of $\alpha$, showing the equilibrium and aging dynamics.

% The selection of an exponential form for $\rho(E)$ allows us to observe glass transition, manifesting as aging dynamics over an infinite time period. While alternate forms for $\rho(E)$ could be considered, such as a Gaussian distribution, these alternatives only exhibit aging behavior for a finite time due to their faster decay of the distribution tail~\cite{monthus1996models}. The simple choice of an exponential $\rho(E)$ successfully captures the basic physics of the glass transition and is thus appropriate for the theoretical study within our model.
% The validity of the chosen form of $\rho(E)$ could potentially be verified experimentally through the measurement of protein trapping times distribution, since an exponential $\rho(E)$ would lead to a power-law distribution of these times~\cite{monthus1996models}. Conducting such an experiment would necessitate high temporal and spatial resolution to precisely monitor the binding and unbinding events of proteins.

To complete the model of an aging protein condensate we propose a constitutive equation of the condensate rheology. Cross-linked elements in the condensate are elastic with a shear modulus $G_0$. When they unbind they can flow with viscosity $\eta_0$. Cross-link binding and unbinding is accounted for by the trap model in Eqs.(\ref{eq:ME2}).
The shear strain rate of an unbound element is $\dot{\epsilon}_u(t)= \sigma(t)/\eta_0$ while the strain of an element in the bound state is $\epsilon_b(t)= \sigma(t)/G_0$, where $\sigma(t)$ is the shear stress.
Assuming the shear stress to be uniform within the condensate, the overall shear strain rate $\dot{{\epsilon}}(t)= P_u(t)\dot{\epsilon}_u(t) + (1 - P_u(t))\dot{\epsilon}_b (t)$ is therefore
\begin{equation} 
\begin{split}
\label{eq:CE}
\dot{\epsilon}(t) = \frac{\sigma(t)}{\eta_0}P_u(t) + \frac{\dot{\sigma}(t)}{G_0}(1 - P_u(t)).
\end{split}
\end{equation}
This is an equation of a viscoelastic Maxwell material with an effective viscosity $\eta_c=\eta_0/P_u(t)$ and an effective elastic modulus $G_c=G_0/(1 - P_u(t))$, that can exhibit aging dynamics described in Eq.(\ref{eq:pulong1}).
In the aging regime $P_u(t)$ decays towards $0$, see Eq.(\ref{eq:pulong1}) and Fig.\ref{Fig:Puplots}b, so that the effective viscosity diverges and the effective elastic modulus decreases towards the value $G_0$.
% Interestingly, a weak decay towards a constant value of the effective elastic modulus, consistent with our model, was indeed observed in the aging protein condensates \cite{jawerth2020protein}. 
For simplicity, in the analytical calculations, we approximate the effective elastic modulus with the value $G_c \simeq G_0$ to which it converges at long times. This approximation is exact at the lowest order in $P_u(t)$, see Appendix \ref{sec:Change_of_Elastic} for details. 
%This is an equation of a viscoelastic Maxwell material with a instantaneous Maxwell time $\nu^{-1}(t)$ where $\nu(t)= \eta_0/(G_0P_u(t))$ that can exhibit aging dynamics described in Eq. (\ref{eq:pulong1}).
%We thus derived the probability of diffusing element to be in the unbound state and the dynamics is qualitatively different depending on the value of $\alpha$.  

%While it is feasible to incorporate the change of elasticity, $G_0(1-P_u(t))$, into the constitutive equation, doing so would result in negligible correction to the equation defined by Eq.(\ref{eq:CE}) in a long time limit. In the equilibrium regime, $P^{eq}_u$ is a constant, meaning that any variation in the elasticity simply translates to a constant shift in elasticity and does not alter the overall behaviour. In the aging regime, $1-P_u(t)$ approximates to $1$, as supported by the numerical values of $P_u(t)$ presented in Fig.\ref{Fig:Puplots}b, rendering it inconsequential to the constitutive equation. This accords with the the minor change of the elasticity during the aging observed in the experiment~\cite{jawerth2020protein} and it is supported by the binding and unbinding kinetics of proteins manifested in $P_u(t)$ in our model. We refer Appendix \ref{sec:Change_of_Elastic} for more detailed discussion.  

\section{\label{sec:CEMG} Active rheology of aging condensates}
%We now write the constitutive equation for aging Maxwell fluid. Because the viscosity arises when the elements are in the unbound, diffusive state, we write the dissipative part of the constitutive equation being proportional to $P_u(t)$. The constitutive equation may be written as,

%where $\bar{\epsilon}(t)$ and $\sigma(t)$ are strain and stress, respectively. $\eta_0$ and $G_0$ are the bare viscosity and elastic modulus, respectively. The bare relaxation rate of Eq.(\ref{eq:CE}) is set by $\nu_0=G_0/\eta_0$. In Eq.(\ref{eq:CE}) we define the time dependent effective viscosity of the condensate, $\eta(t)\equiv \eta_0/P_u(t)$, in turn, the time dependent relaxation rate of aging Maxwell fluid, $\nu(t)\equiv\nu_0 P_u(t)$. 
\subsection{\label{sec:CEMGa} Relaxation function of a Maxwell glass}
We now derive and discuss the linear response of a viscoelastic material described by Eqs.(\ref{eq:ME2}) and (\ref{eq:CE}) with a constant elastic modulus $G_0$. % Here, Response of a viscoelastic material to an imposed strain rate is characterized by the relaxation function $K(t, t')$, which describes the response of the stress to the strain rate (Appendix.\ref{sec:rheology}). 
In order to compare our model with rheology experiments, we solve Eq.(\ref{eq:CE}) for the shear stress
% \begin{equation} 
% \begin{split}
% \label{eq:sigmasol}
% \sigma(t) = e^{-\int_{0}^{t}dt'\nu(t') }\int_{0}^{t} dt' G_0 \dot{\epsilon}(t')e^{\int_{0}^{t'} dt'' \nu(t'')},
% \end{split}
% \end{equation}
\begin{equation} 
\begin{split}
\label{eq:sigmasol}
\sigma(t) = \int_{0}^{t} dt' K(t,t')\dot{\epsilon}(t'),
\end{split}
\end{equation}
where
\begin{equation} 
\begin{split}
\label{eq:Ktt'nu}
K(t,t') = G_0 e^{-\frac{G_0}{\eta_0}\int_{t'}^{t} dt'' P_u(t'')} 
\end{split}
\end{equation}
is the relaxation function and $t= 0$ corresponds to the sample preparation time at which $\sigma(0)= 0$. %Here, we denote the instantaneous rate of relaxation $\nu(t)=P_u(t)G_0/\eta_0 $.

For $\alpha > 1$, the equilibrium steady state $P^\text{eq}_u(t)$ exists and the relaxation function becomes $K(t-t') = G_0\exp\big(- P^\text{eq}_uG_0 /\eta_0   \cdot (t-t')\big)$. This is the exponential relaxation with the rate $ P^\text{eq}_uG_0/\eta_0  $, which corresponds to a Maxwell fluid.
For $0 < \alpha < 1$, no steady state exists, and the relaxation function exhibits glassy behavior. In the asymptotic regime, $P_u(t)$ follows Eq.(3), from which we obtain:
\begin{equation} 
\begin{split}
\label{eq:Ktt'}
K(t,t') \simeq G_0 \exp \Big[{-\frac{\kappa G_0}{\alpha \Gamma_0 \eta_0}\big((\Gamma_0 t)^\alpha - (\Gamma_0 t')^\alpha \big)}\Big].
\end{split}
\end{equation}
Therefore in the aging regime, the relaxation function takes the form of a stretched exponential that often appears in the relaxation of glass forming materials~\cite{wuttke1996structural,phillips1996stretched}. Note that the time translational invariance is broken in Eq.(\ref{eq:Ktt'}), a signature of the aging regime. We refer to the relaxation function in Eq.(\ref{eq:Ktt'}) as the relaxation function of an aging Maxwell fluid, i.e., Maxwell glass.
\subsection{Age dependent relaxation time}
We consider an experimental protocol where the system is prepared at $t= 0$ and the system is strained  starting at the waiting time $t_w$. The resulting stress is written as
\begin{equation} 
\begin{split}
\label{eq:1/2}
\sigma(t) \simeq \int_{t_w}^{t} dt' K(t,t')\dot{\epsilon}(t'),
\end{split}
\end{equation}
where $\epsilon(t_w)=0$. 
We consider the relaxation function in terms of the observation time $\tau = t - t_w$. 
%We introduce the change of variables $t=t_w+\tau$ and $t'=t_w+\tau'$:
%\begin{equation} 
%\begin{split}
%\label{eq:sigKtw}
%\sigma(t_w+\tau) =& \int_{0}^{\tau} d\tau' K(t_w+\tau,t_w+\tau')\dot{\epsilon}(t_w+\tau').
%\end{split}
%\end{equation}
In the limit of a short observation time compared to the waiting time $\tau \ll t_w$, %, the exponent of $K(t_w+\tau,t_w+\tau')$ %in Eq.(\ref{eq:sigKtw}) 
%can be expanded up to the first order of $\tau/t_w$ and $\tau'/t_w$, and we find % Consequently, we obtain the constitutive equation for $\tau/t_w \ll 1$,
% \begin{equation} 
% \begin{split}
% \label{eq:Ktwtau}
% K(t_w + \tau,t_w + \tau') \simeq G_0 e^{-\kappa \nu_0 (\Gamma_0t_w)^{\alpha-1}(\tau-\tau')}.
% \end{split}
% \end{equation}
%\begin{equation} 
%\begin{split}
%\label{eq:Ktw}
%\sigma_{t_w}(\tau) \simeq \int_0^\tau K_{t_w}(\tau-\tau') \dot{\epsilon}_{t_w}(\tau'),
%\end{split}
%\end{equation}
the relaxation function $K(t_w + \tau, t_w + \tau')$ can be approximated by a time translation invariant function
\begin{equation} 
\begin{split}
\label{eq:Ktwtautau'}
K_{t_w}(\tau-\tau') \equiv G_0 e^{-\frac{\kappa G_0}{\eta_0} (\Gamma_0t_w)^{\alpha-1}(\tau-\tau')}.
\end{split}
\end{equation}
%where $\sigma_{t_w}(\tau) \equiv \sigma(t_w+\tau)$ and $\epsilon_{t_w}(\tau) \equiv \epsilon(t_w+\tau)$.
This relaxation function shows that a Maxwell glass behaves as a Maxwell fluid when observed on short times $\tau \ll t_w$, but with age-dependent relaxation time
\begin{equation} 
\begin{split}
\label{eq:tauc}
\tau_c(t_w) = \frac{\eta_0}{\kappa G_0}(\Gamma_0 t_w)^{1-\alpha}.
\end{split}
\end{equation}
% This result provides a connection between aging in an underlying cross-linker network and the experimentally observed  age-dependent relaxation time reported in the protein condensates rheology experiments~\cite{jawerth2020protein}. 

The age-dependent Maxwell relaxation time derived here provides a connection between underlying dynamics of cross-linker network and Maxwell glass rheology~\cite{jawerth2020protein}. The aging of the Maxwell relaxation time stems from the stretched exponential relaxation in Eq.(7) that reflects the glassy nature of the material.

% Eq.(\ref{eq:ME2}) and Eq.(\ref{eq:CE}) define the model for the rheology of condensates in our study. That is, Eq.(\ref{eq:ME2}) describes the dynamics of $P_u(t)$ and Eq.(\ref{eq:CE}) defines the relation between stress and strain by using $P_u(t)$ obtained from Eq.(\ref{eq:ME2}). We are now to examine the rheology of our model using Active rheology.

\subsection{\label{sec:AC} Instantaneous complex modulus}
The relaxation time $\tau_c$ in a Maxwell fluid is related to the complex modulus as $G(\omega)= i \omega \tau_c G_0/(1 + i \omega \tau_c)$~\cite{furst2017microrheology}. The complex modulus $G(\omega)= G^\prime(\omega) + iG^{\prime \prime}(\omega)$, where $G^\prime(\omega)$ and $G^{\prime \prime}(\omega)$ represent the storage and loss moduli, respectively, characterizes the linear response of a time-translation-invariant material as a function of the angular frequency $\omega$. However, for an aging material, $G(\omega)$ is not a well-defined observable. Nevertheless, a frequency-dependent linear response can still be employed if the observation time window $\tau$ is short enough such that the material properties do not undergo significant changes during the observation (Appendix \ref{sec:rheology}).
%The value of $\tau$ is restricted by the aging of the material, as changes in the material properties during observation are not taken into account. 
To remove the restriction of a short observation time window, which limits the applicability of active rheology for aging material, we now introduce an analytic signal method that allows us to define the instantaneous complex modulus of an aging material $G(\omega, t, t_w)$ at time $t$ and at frequency $\omega$, similar to the time-varying viscoelastic spectrum \cite{fielding2000aging}, see Appendix~\ref{sec:Hilbert}.

%In the method of active rheology, external modulation, typically weak sinusoidal force or displacement having frequency $\omega$, is applied to the probing materials. One can compute the storage and loss modulus of the material from the relation between the input and output (either stress or strain) as explained in Appendix \ref{sec:rheology}. The storage and loss modulus defined in this manner is only applicable for the stationary stress ($\sigma$) and strain ($\epsilon$), giving constant change of the amplitude ($\sigma/\epsilon$) and phase shift ($\delta \varphi$). This procedure to characterize the material property is not generally applicable to the aging materials where the material properties slowly change over time. Because of this restriction, active rheology experiments must be conducted within a time scale sufficiently shorter than the change of the material properties. To overcome this restriction, we generalize the storage and loss modulus using the instantaneous amplitude and phase of the signal, defining the instantaneous storage and loss modulus. The instantaneous storage and loss modulus enable us to extract the instantaneous material property, resolving the limitation. Phase and amplitude extraction of time varying signals is a common task in signal processing~\cite{cohen1995time}. 
The analytic signal of a function $f(t)$
is defined as $f_a(t) \equiv f(t) +i \mathcal{H}[f(t)](t)$, where $\mathcal{H}$ is the Hilbert transform, see Appendix~\ref{sec:Hilbert}. The analytic signal $f_a(t)$ is a complex function and can be written in the polar form, $f_a(t) = |f_a(t)| \exp(i\varphi(t))$, where $|f_a(t)|$ is the instantaneous amplitude, also called envelope, and $\varphi(t)=\arg[f_a(t)]$ is the instantaneous phase of the signal $f(t)$. 
%When calculating the analytic signal of experimental or numerical data we set $f(t)= 0$ outside of the observation time window.
%(Appendix~\ref{sec:Hilbert}) to extract the instantaneous amplitude and phase of the input and output signals. Analytic signal for a signal $f(t)$ 
% For an imposed sinusoidal shear strain with frequency $\omega$  starting at $t=t_w$, $\bar{\epsilon}(\omega,t,t_w)=\Theta(t-t_w)\epsilon(t)$ where $\epsilon(t)=\epsilon_0 \cos{(\omega t+\varphi)}$ and $\Theta$ is the Heaviside step function.
Using this definition of the analytic signal, we define the instantaneous complex modulus as 
% using the analytic signal of the imposed shear strain and measured shear stress $\sigma(\omega,t,t_w)$, as 
\begin{equation} 
\begin{split}
\label{eq:Gtdef}
G(\omega,t,t_w) \equiv & \frac{\sigma_a(\omega,t,t_w)}{\epsilon_a(\omega,t)}\\ =&\frac{|\sigma_a(\omega,t,t_w)|}{|\epsilon_a(\omega,t)|}\exp \big(i\delta \varphi(\omega,t,t_w)\big),
\end{split}
\end{equation}
where $\delta \varphi(\omega,t,t_w)$ is the instantaneous phase difference between shear strain  and stress.
Here $\sigma_a(\omega,t,t_w)$ is the analytic signal of measured shear stress $\sigma(\omega,t,t_w)$ in response to an imposed sinusoidal shear strain  $\bar{\epsilon}(\omega,t,t_w)=\Theta(t-t_w)\epsilon(\omega,t)$ with frequency $\omega$ starting at $t=t_w$, where $\epsilon(\omega,t)=\epsilon_0 \cos{(\omega t+\varphi_0)}$ and $\Theta$ is the Heaviside step function. $\epsilon_0$ and $\varphi_0$ are the amplitude and initial phase of the shear strain, respectively. The analytical signal of the strain is $\epsilon_a(\omega,t)=\epsilon_0e^{i(\omega t + \varphi_0)}$.  
%The $\omega$ dependence of $\sigma_a(\omega,t)$ and $\epsilon_a(\omega,t)$ enters via the frequency of the input signal. 
%When the sinusoidal input signal starts at $t=t_w$, the dependence on $t_w$ may be included (Appendix~\ref{sec:Hilbert}). 
%Since the instantaneous complex modulus does not require short time observations for different $t_w$, it is natural to fix $t_w=0$ and follow the evolution of $G(\omega,t)$ as a function of only $t$ for a given $\omega$. 
The instantaneous complex modulus $G(\omega,t,t_w)$ is a generalization of the conventional complex modulus $G(\omega)$ to the time dependent signals and they become equal for a time translation invariant system, see Appendix~\ref{sec:Hilbert}. It reduces to the time-varying viscoelastic spectrum defined in Ref.\cite{fielding2000aging} for slow aging limit as discussed in Appendix~\ref{sec:Hilbert}.  
%It is similar to the  time-varying viscoelastic spectrum defined in Ref.\cite{fielding2000aging}, as discussed in Appendix~\ref{sec:Hilbert}. %Furthermore, the conventional complex modulus of Maxwell fluid  is recovered from Eq.(\ref{eq:Gtdef}) when viscosity does not change in time [$P_u(t)=\text{const.}$ in Eq.(\ref{eq:CE})], see Appendix~\ref{sec:Hilbert}.

\begin{figure}
\begin{center}
%\scalebox{1.5}[1.5]{
\includegraphics[width=0.5\textwidth]{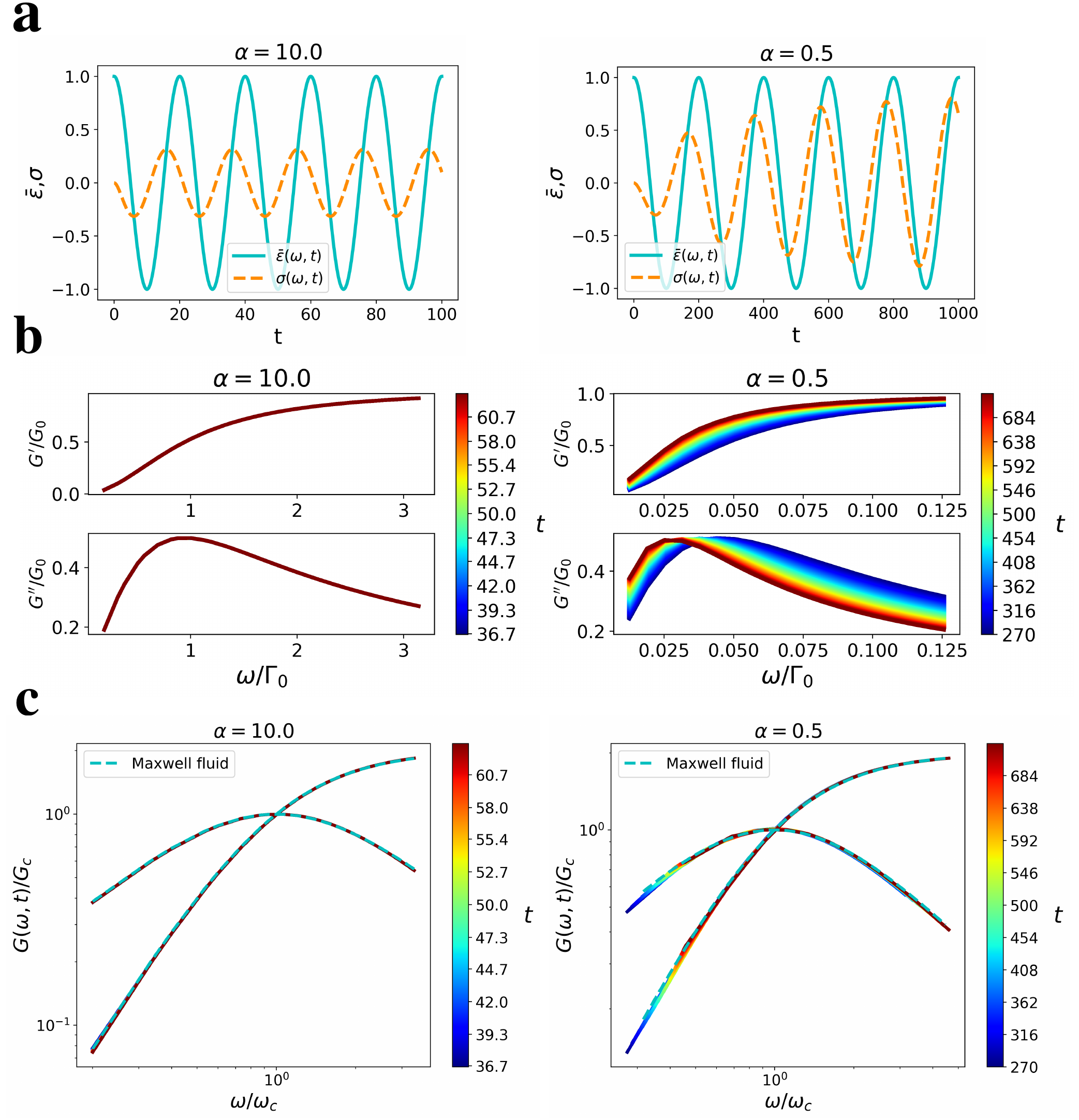}
%}
\end{center}
\caption{\label{Fig:AR} Active rheology for the Maxwell fluid and glass. In the case of $\alpha=10.0$ the system has a stationary equilibrium state and thus behave as conventional Maxwell fluid. For $\alpha=0.5$, the system shows aging, thus behaving as the Maxwell glass. The unit time is $1/\Gamma_0$ in Eq.(\ref{eq:ME2}). {\bf (a)} The input shear strain $\bar{\epsilon}(\omega,t)$ (cyan solid line) and the output shear stress $\sigma(\omega,t)$  (orange dashed lines). $\omega =\pi/10$ for $\alpha=10.0$ and $\omega =\pi/100$ for $\alpha=0.5$. {\bf (b)} The instantaneous complex modulus $G(\omega,t)$ in equilibrium and aging regime. The real and imaginary part of $G(\omega,t)$ is $G^\prime(\omega,t)$ and $G^{\prime \prime}(\omega,t)$, respectively. {\bf (c)} The collapse of the $G(\omega,t)$ for different instances onto the single master curve of the Maxwell fluid (dashed line in cyan). The bare viscosity is set to $\eta_0=0.5$. We fix $\beta_0$ to $1$ and vary $\beta$. Detailed numerical procedures are in Appendix~\ref{sec:NAR}. }
\end{figure}

We use the instantaneous complex modulus to analyze the rheology of our model. For simplicity we choose a waiting time $t_w=0$, which does not affect aging process in our model. We therefore omit the $t_w$ dependence in the following. We solve  Eq.(\ref{eq:CE}) with Eq.(\ref{eq:ME2}) numerically for the sinusoidal shear strain as input $\bar{\epsilon}(\omega,t)$ and obtain the shear stress $\sigma(\omega,t)$ as output. Fig.\ref{Fig:AR}a shows the shear strain and stress for  $\omega=\pi/10$ and $\omega=\pi/100$ for $\alpha=10$ and $\alpha=0.5$, respectively. For $\alpha=10$, the strain is stationary, reflecting the equilibrium  viscosity in Eq.(\ref{eq:CE}). In contrast, for $\alpha = 0.5$ the amplitude of shear stress increases in time due to aging, reflected in changing viscosity $\eta_0/P_u(t)$. In Fig.\ref{Fig:AR}b, we calculate the real and imaginary part of the instantaneous complex modulus, $G^\prime(\omega,t)$ and $G^{\prime \prime}(\omega,t)$, respectively, for a range of input frequencies. For $\alpha=10$,  $G(\omega,t)$ does not depend on the time. On the contrary, we observe a striking difference for $\alpha=0.5$: the instantaneous complex modulus shifts to lower frequencies over time, showing that the characteristic relaxation time of the material increases, as shown in Fig.\ref{Fig:AR}b, right panel. Such aging behavior was observed experimentally in the protein condensates~\cite{jawerth2020protein}.
Moreover, Jawerth {\it et al.}~\cite{jawerth2020protein} demonstrated that experimentally measured complex moduli in the Maxwell glass collapse when rescaled by $G_c$ and frequencies by $\omega_c$, where $G_c$ and $\omega_c$ are defined by $G^\prime(\omega_c,t)=G^{\prime \prime}(\omega_c,t)= G_c$. We show in Fig.\ref{Fig:AR}c that our numerically evaluated complex moduli indeed collapse on a single master curve of the Maxwell fluid when rescaled moduli and frequency by $G_c$ and $\omega_c$, respectively.

% Instantaneous complex moduli $G(\omega,t)$ at different times collapse when their magnitudes are rescaled by $G_c$ and frequencies by $\omega_c$, where $G_c$ and $\omega_c$ are defined by $G^\prime(\omega_c,t)=G^{\prime \prime}(\omega_c,t)= G_c$. We show in Fig.\ref{Fig:AR}c that our numerically evaluated complex moduli indeed collapse on a single master curve of the Maxwell fluid  as was observed for experimentally measured complex moduli in the Maxwell glass~\cite{jawerth2020protein}. 

% In concluding the section on active rheology, we analyzed the time-dependent material properties of the model [Eq.(\ref{eq:CE})] using instantaneous complex modulus. Instantaneous complex modulus does not impose the limitation to the observation time $\tau$. 
% Nevertheless, we found the curves of the complex modulus for different times collapse accurately on the master curve of Maxwell fluid. This implicates the high reliability of the approximation in the relaxation function [Eq.(\ref{eq:Ktwtautau'})] and support the experimental procedure commonly employed in Active rheology experiments using short observation time $\tau$.   
 
\section{\label{sec:PR} Fluctuation-response relations in aging condensates}
In an equilibrium system, the relaxation of spontaneous fluctuations and the linear response to an external perturbation are closely related by the fluctuation-dissipation theorem~\cite{kubo2012statistical}. Using the generalized Stokes-Einstein relation derived from the fluctuation-dissipation theorem, rheological properties of the material can be determined from equilibirum fluctuations~\cite{mason2000estimating, jawerth2020protein}. %   is governed by the same linear relaxation law as  in an equilibrium system In passive rheology, one tracks the trajectories of diffusing elements. The fluctuation of trajectories, such as MSD, encodes the material properties in equilibrium~\cite{mason2000estimating}.
Although the equilibrium fluctuation-response relations do not apply in the aging materials, we derive specific fluctuation-response relations that characterise the aging Maxwell fluid.

To this end, we consider a spatially resolved version of  Eq.(\ref{eq:ME2}) that takes into account diffusion of unbound elements
\begin{subequations}
\label{eq:ME1}
  \begin{align} 
  \label{eq:ME1a}
   \frac{1}{\Gamma_0}\frac{\partial  p_b(x,E,t)}{\partial t}=&-p_b(x,E,t) e^{-\beta E}+p_u(x,t)\rho(E) ,\\
   \label{eq:ME1b}
   \frac{1}{\Gamma_0}\frac{\partial p_u(x,t)}{\partial t}  =&\frac{D_0}{\Gamma_0}  \frac{\partial^2  p_u(x,t)}{\partial x^2}- p_u(x,t)  \nonumber \\
   & + \int_0^{\infty}dE p_b(x,E,t)e^{-\beta E},
  \end{align}
\end{subequations}
with the initial condition $p_b(x,E,0)= 0$ and $p_u(x,0)= \delta(x)$. In Eq.(\ref{eq:ME1}), $p_b(x,E,t)$ is the probability density of elements bound at position $x$ with energy $E$ at time $t$ and $p_u(x,t)$ is the density of diffusing elements at position $x$ at time $t$. %It can be shown that integrating Eq.(\ref{eq:ME1}) over the coordinate $x$ reduces to Eq.(\ref{eq:ME2}), assuming $p_u(x,t)$ vanishes at infinity.

% where $x$ is a component of the vector $\vec{r}$, $\Gamma_0$ is the bare rate setting the time scale of Eq.(\ref{eq:ME1}), and $D_0$ is the bare diffusion coefficient of the DP. We assume that the condensate is isotropic, thus the evolution of the probability density in $x,y,z$ direction independently follows Eq.(\ref{eq:ME1}). 
% The physical justification may be given as expecting the probability of the formation of cross-links to obey a Poisson process, $P(n)\sim e^{-n\bar{E}/E_0}$, where $n$ is the number of cross-links, $\bar{E}$ is the binding energy for one chemical cross-link, and $E_0$ is a characteristic energy scale. 

% The bound state obstructs the diffusion of the elements while they freely diffuse in the unbound state. Thus, we derive the effective diffusion coefficient of DP using Eq.(\ref{eq:ME1}). 
The mean square displacement of fluctuating elements is 
\begin{equation} 
\begin{split}
\label{eq:MSDdef}
\langle \Delta x^2 \rangle(t)=\Delta_u(t)  + \int_0^\infty dE \Delta_b (E,t),
\end{split}
\end{equation}
where we have defined the positional variance of diffusing and bound states, respectively, as 
\begin{equation} 
\begin{split}
\label{eq:MSDeach}
&\Delta_u (t)\equiv \int_{-\infty}^\infty dx x^2 p_u(x,t); \\ 
&\Delta_b(E,t) \equiv  \int_{-\infty}^{\infty} dx x^2 p_b(x,E,t).
\end{split}
\end{equation}
Using Eqs.(\ref{eq:ME1}) and Eqs.(\ref{eq:MSDeach}), we obtain the time evolution of the mean squared displacement, 
\begin{subequations}
\label{eq:}
\label{eq:MSDME}
  \begin{align} 
  \label{eq:}
   \frac{1}{\Gamma_0}\frac{\partial \Delta_b(E,t)}{\partial t}&= -\Delta_b(E,t) e^{-\beta E}+\Delta_u (t)\rho(E) ,\\
   \label{eq:}
   \frac{1}{\Gamma_0}\frac{\partial\Delta_u (t)}{\partial t}  &= 2\frac{D_0}{\Gamma_0}P_u(t) - \Delta_u (t) \nonumber \\
   &+ \int_0^{\infty}dE \Delta_b(E,t)e^{-\beta E},
  \end{align}
\end{subequations}
with the definition,
\begin{equation} 
\begin{split}
\label{eq:}
P_u(t) \equiv \int_{-\infty}^{\infty} dx p_u(x, t).
\end{split}
\end{equation}
The expression for the effective diffusion coefficient, $D(t)$, can be obtained by taking the time derivative of Eq.(\ref{eq:MSDdef}) and using Eq.(\ref{eq:MSDME}),
\begin{equation} 
\begin{split}
\label{eq:D(t)}
\frac{d}{dt} \langle \Delta x^2 \rangle(t) = 2 D_0P_u(t),  \\ 
\end{split}
\end{equation}
leading to
\begin{equation} 
\begin{split}
\label{eq:}
D(t) \equiv D_0 P_u(t).  \\ 
\end{split}
\end{equation}
%Since $P_u(t)$ is the probability of the diffusing element being in an unbound state, 
Eq.(\ref{eq:D(t)}) states that the effective diffusion coefficient is proportional to the probability that the element being in the diffusive state. 
%Observing the tracer particle after waiting time $t_w$ over short time, $\tau \ll t_w$, we find the waiting time dependent diffusion coefficient in the aging regime 
%\begin{equation} 
%\begin{split}
%\label{eq:D(tw)}
%D(t_w) \simeq D_0 \kappa (\Gamma_0 t_w)^{\alpha-1}, \\ 
%\end{split}
%\end{equation}
%which accounts for the time dependent diffusion coefficient observed in experiments~\cite{jawerth2020protein}. 
%Note that $\langle \Delta x^2 \rangle(t)$ includes a constant offset, $2k_BT/G_0$, which stems from short time elastic relaxation in the trap (Appendix \ref{sec:FRR}). 

We now obtain a relation between the aging relaxation function 
and the mean squared displacement at different times using Eq.(\ref{eq:Ktt'nu}) and Eq.(\ref{eq:D(t)})
\begin{equation} 
\begin{split}
\label{eq:Kx2}
K(t,t') = G_0\exp\Big(-\frac{G_0}{2D_0\eta_0}\big(\langle \Delta x^2 \rangle(t)-\langle \Delta x^2 \rangle(t')\big)\Big). \\ 
\end{split}
\end{equation}
This exact relation connects the time dependent rheology $K(t, t')$ of the Maxwell glass to the passive rheology characterised by the mean squared displacement $\langle \Delta x^2 \rangle (t)$. %We note that the observation time of the MSD can be long or short, and for short time intervals   ($\tau \ll t_w$), Eq.(\ref{eq:Kx2}) reduces to a Maxwell fluid with properties that depend on waiting time $t_w$.    
Alternatively, we can write the second relation between mean squared displacement and linear response function.  Using the strain-stress response function $\chi(t,t')$ defined as
\begin{equation} 
\begin{split}
\label{eq:}
\epsilon(t)=\int_0^{t} \chi(t,t')\sigma(t'),
\end{split}
\end{equation}
we obtain (see Appendix \ref{sec:FRR})
\begin{equation} 
\begin{split}
\label{eq:GFDT}
\Theta(t-t')\frac{d}{dt'}\langle \Delta x^2(t') \rangle = 2k_BT \chi(t,t').
\end{split}
\end{equation}
Eq.(\ref{eq:GFDT}) stems from the fact that both the time dependence of the diffusion coefficient $D(t)$ and of the active response given in Eq.(\ref{eq:CE}) are governed by $P_u(t)$. We have used $D_0=k_BT/\eta_0$ implying that diffusion coefficient of the unbound elements satisfies the Einstein relation. Note that Eq.(\ref{eq:GFDT}) is similar to but different from the time translation invariant fluctuation dissipation theorem in equilibrium. It applies to the aging Maxwell model and has both $t$ and $t'$ dependence signifying the glassy behavior.

\section{Discussion}
%{\it Aging exponent.}

We have presented a mean-field model of aging biological condensates, based on a minimal trap model that exhibits  glassy behaviour~\cite{bouchaud1992weak}. Our model recapitulates aging rheology recently observed in biological condensates termed Maxwell glass.  
A Maxwell glass exhibits at all times Maxwell fluid behaviour with an age-dependent relaxation time, corresponsingly the viscosity is age dependent and diverges for long times, even though the system remains fluid. In addition, it was observed that the elastic modulus decreased slightly but remained roughly constant~\cite{jawerth2020protein}. Interestingly the complex modulus  measured at different ages collapses on master curves describing a Maxwell fluid. 
In the aging regime of our model %The constitutive equation of our model, Eq.(\ref{eq:CE}), contains the dynamics of the trap model given in Eq.(\ref{eq:ME2}). T
the fraction of unbound elements decays with time as a power law $P_u(t)\sim t^{\alpha-1}$ ($\alpha < 1$). This leads to a diverging effective viscosity $\eta_0/P_u(t)$ and a weakly decreasing effective modulus $G_0/(1-P_u(t))$ that approaches a finite value.  The relaxation function $K(t,t')$ in our model exhibits a stretched exponential that decays at low temperatures, a characteristic for glassy systems. The resulting Maxwell relaxation time is age dependent and increases with waiting time $t_w$ as a power law $\tau_c \sim t_w^{1 - \alpha}$. The complex modulus determined in our model collapses on curves describing a Maxwell model, consistent with experiment.

For such an aging material for which time translation invariance is not obeyed,
 defining the frequency dependent complex modulus poses a challenge. To
 overcome this challenge, we introduce the time-dependent instantaneous complex modulus as a generalization of the conventional complex modulus at steady state. The instantaneous complex modulus is based on analytic signal construction and remains well-defined even in non-stationary systems where approximative measures of the conventional complex modulus would fail.

Our theory is a phenomenological mean-field model that  captures key characteristic rheological properties of protein condensates ~\cite{jawerth2020protein}.  Different future extensions of our study will be of interest.  These include a microscopic model of the protein condensate network, for example by building on models for dynamic cross-linked networks such as Flory's addition-substraction network theory~\cite{flory1961thermodynamic,fricker1973effects} and transient network theory~\cite{tanaka1992viscoelastic,tanaka1992viscoelastic_2}. Moreover, another interesting extension would be to consider the coupling between externally applied shear stress and the unbinding rate of cross-linked proteins. This could potentially provide insight into plastic events, a phenomenon that has been investigated within the context of amorphous materials~\cite{hebraud1998mode,nicolas2018deformation} and particularly in connection to aging~\cite{sollich2017aging}. 

A power-law dependence of the relaxation time on the waiting time has been observed in different system. The aging exponent $\mu$, which describes the growth of relaxation time with waiting time as $\tau_c\sim t_w^\mu$ has been introduced in the seminal work~\cite{struik1977physical}. 
%\begin{equation} 
%\begin{split}
%\label{tauStruik}
%\tau_{c}(t_w)\sim (t_w)^{\mu}.
%\end{split}
%\end{equation}
In many polymeric materials, the relaxation time grows sublinearly, $\mu \simeq 0.5 - 1$ ~\cite{berthier2011theoretical}. 
In our model $\mu=1-\alpha$ [see Eq.(\ref{eq:tauc})] and in the aging regime with $0<\alpha<1$, we find a sublinear dependence of $\tau_c$ on $t_w$ for a Maxwell glass, consistent with the sublinear behavior seen in many experiments on non biological materials.

Interestingly, recent experiments suggest that $\mu$ could be larger than $1$ in  protein condensates. For example, for the PGL-3 protein, $\mu\simeq 6.4$ and $\mu\simeq 2.1$ were estimated for different salt conditions (150 mM KCl and 100mM KCl, respectively)~\cite{jawerth2020protein}.  
Our current model does not account for such high values of $\mu$, as they would require negative values of $\alpha$ and we currently do not have an explanation of this discrepancy. %One possibility is that it stems from the simple choice of
%an exponential distribution of trap strength in our model, which may not capture the trap statistic in a condensate.
There are only very few other systems where $\mu >1$ was measured. An example is polycarbonate (see for instance Fig.15 in \cite{struik1977physical}). Further research will be required to find out whether $\mu >1$  is a robust feature of biological protein condensates, and if so, what is the origin of such a different behavior
in comparison to aging of non-biological polymers. 
One possible explanation of the rapid aging observed in protein condensates, is that the system may not yet be exploring the tail of the distribution $\rho(E)$ for large $E$ within the experimental time-scales. Instead, the system may be exploring smaller $E$, where the distribution $\rho(E)$ might not be a decreasing function of $E$. This could lead transiently to a relaxation time that grows exponentially with age.
The functional form of the distribution $\rho(E)$ could be probed experimentally, for example, through a measurement of the distribution of protein trapping times.

Finally, we have obtained an exact relation  between the relaxation function and the mean squared displacement of particles in the aging regime (Eq. (\ref{eq:Kx2})). This relation is similar to the fluctuation-dissipation theorem that holds for equilibrium systems but it applies to the out-of-equilibrium Maxwell glass.
In out-of-equilibrium aging systems, the generalized fluctuation-dissipation theorem has been hypothesized and verified for various models, resulting in the definition of an effective temperature~\cite{cugliandolo1997energy,cugliandolo2011effective,berthier2001glassy}. 
The fluctuation-response relation, given by Eq.(\ref{eq:GFDT}), does not require an effective temperature. Instead, it directly connects the response function to the fluctuations observed in Maxwell glass. %This relation shows how rheological properties of aging protein condensates can be inferred from microrheological experiments. %and provides a framework to interpret recent microrheology experiments to study aging protein condensates. 

\appendix

\section {\label{sec:rheology} Rheology of glassy materials}
%In this introductory section, we introduce several essential concepts underlying in our theory.  
%We illustrate the connection of the theory of linear viscoelasticity to the experimental outcomes using microrheology. 

Soft materials, including protein condensates, behave as viscoelastic fluids. We consider a material that was prepared at $t=0$ and start measuring the material properties after a waiting time, $t=t_w$. Linear viscoelasticity is characterized by the linear constitutive relation between stress ($\sigma$) and strain (${\epsilon}$). We consider the stress and strain relative to $t=0$, which subsume the effect of stress and strain at $t=0$ into $\sigma(t)$ and ${\epsilon}(t)$, respectively. The linear constitutive relation reads
\begin{equation} 
\begin{split}
\label{eq:LVE_G}
\sigma(t) = \int_0^{t} G(t,t'){\epsilon}(t')dt' \quad ,  \\ 
\end{split}
\end{equation}
where we consider a general case without time translation symmetry~\cite{fielding2000aging}. 
Here, $G(t,t')$ is dynamic modulus determining the linear relation between the shear strain and stress. %The dynamic modulus stands on the solid description that the stress is proportional to the strain of the material. 
We can alternatively write the relation between stress and strain-rate, 
\begin{equation} 
\begin{split}
\label{eq:LVE_K}
\sigma(t) = \int_0^{t} K(t,t')\dot{{\epsilon}}(t')dt',  \\ 
\end{split}
\end{equation}
where $\dot{{\epsilon}}$ is the rate of deformation. $K(t,t')$ is called relaxation function. 
%The relaxation function is based on the fluid description: the stress is proportional to the strain-rate. 
We obtain the relation between $G(t,t')$ and $K(t,t')$ by applying partial integration to Eq.(\ref{eq:LVE_K}),
\begin{equation} 
\begin{split}
\label{eq:K}
%G(t,t') = -\frac{dK(t,t')}{dt'} + 2\big(\delta(t-t')-\delta(t')\big)K(t,t').\\ 
G(t,t') = -\frac{dK(t,t')}{dt'} + 2\delta(t-t')K(t,t').\\ 
\end{split}
\end{equation}
The factor $2$ in the above relation is to account for the delta function integrated at the boundary. We used the fact that ${\epsilon}(0)=0$. 
We can also write the linear relationship between stress and strain using the response function, $\chi(t,t')$,
\begin{equation} 
\begin{split}
\label{eq:LVE_chi}
{\epsilon} (t) = \int_0^{t} \chi(t,t')\sigma(t')dt' \quad.  \\ 
\end{split}
\end{equation}

%We may choose either of the two descriptions depending on the interest. 
% For active rheology (Fig.\ref{Fig:schematics}a; left), it is common to use the dynamic modulus, because one can directly compute the modulus using oscillatory signals. The relaxation function is a natural description for passive rheology (Fig.\ref{Fig:schematics}a; right) because tracer elements diffusing in the material are influenced by the viscous damping, which is proportional to the velocity of the tracer. 

% By the physical requirement, $G(t,t')=0$ and $K(t,t')=0$ for $t<t'$ (causality) and $t'<0$ (no response before the preparation), we can extend the limit of the integral in Eq.(\ref{eq:LVE_G}) and Eq.(\ref{eq:LVE_K}) to $\pm \infty$. 
When the probing material is in thermodynamic equilibrium and independent on initial conditions, above response functions depend only on the time interval $t-t'$: $G(t-t')$, $K(t-t')$, and $\chi(t-t')$, corresponding to the time translational invariance. Time translational invariance allows us to apply the convolution theorem for the Laplace transform to Eq.(\ref{eq:LVE_G})-(\ref{eq:LVE_chi}), leading to the simple expressions:
\begin{equation} 
\begin{split}
\label{}
\sigma(s) = G(s) {\epsilon}(s);
\end{split}
\end{equation}
\begin{equation} 
\begin{split}
\label{}
%\sigma(s) = sK(s) \bar{\epsilon}(s)-K(s)\bar{\epsilon}(0),
\sigma(s) = sK(s) {\epsilon}(s);
\end{split}
\end{equation}
and 
\begin{equation} 
\begin{split}
\label{}
%\sigma(s) = sK(s) \bar{\epsilon}(s)-K(s)\bar{\epsilon}(0),
{\epsilon}(s) = \chi(s) \sigma(s).
\end{split}
\end{equation}
We specified the quantities in the Laplace space by the argument $s$.  We use same convention to denote the quantities in Laplace space $(s)$ and in Fourier space ($\omega$). Therefore the response functions have relation $G(s) = s K(s)=1/\chi(s)$ when time translational invariance is satisfied. For causal functions, such as $G(t,t')$, $K(t,t')$, and $\chi(t,t')$, the Fourier transform is readily obtained from the Laplace transform, by analytic continuation: $s \rightarrow i \omega$. Thus, the analytic continuation may give the equivalent relation in the Fourier space, $G(\omega) = i\omega K(\omega)=1/\chi(\omega)$.   

The dynamic modulus in  Fourier space $G(\omega)$, is often referred to as complex modulus~\cite{chen2010rheology}: 
\begin{equation} 
\begin{split}
\label{}
G(\omega) = G^\prime(\omega) + i G^{\prime \prime}(\omega),
\end{split}
\end{equation}
where the real part $G^\prime(\omega)$ is the storage modulus, and the imaginary part $G^{\prime \prime}(\omega)$ is the loss modulus. The storage modulus and the loss modulus reflect the elastic and viscous component of the material response, respectively. The moduli $G^\prime(\omega)$ and $G^{\prime \prime}(\omega)$ may be obtained using active rheology. Depending on the experimental setup, we can choose either strain or stress as input and output signal. Here, we choose, strain as the input and stress as the output. Using a sinusoidal input strain with frequency $\omega$, and amplitude ${\epsilon}(\omega)$, one can determine the moduli by measuring the steady-state output stress, $\sigma(\omega)$, from the amplitude change and the phase shift: 
\begin{subequations}
\label{eq:}
  \begin{align} 
  \label{eq:}
   G^\prime(\omega)&=\frac{\sigma(\omega)}{{\epsilon}(\omega)}\cos(\delta \varphi(\omega));\\
   \label{eq:}
   G^{\prime \prime}(\omega)&=\frac{\sigma(\omega)}{{\epsilon}(\omega)}\sin(\delta \varphi(\omega)),
  \end{align}
\end{subequations}
where $\delta \varphi$ is the phase difference between input and output sinusoidal signals.

In contrast to a material at thermodynamic equilibrium, glassy material, on the other hand, violates time translational invariance due to the slow relaxation which implies that memory about the initial state is not lost. The consequence is the explicit dependence on the two time scales in the complex modulus and the relaxation function, $G(t,t')$ and $K(t,t')$. We introduce the waiting time ($t_w$), the time between the preparation of the material ($t=0$) and the start of the measurement, and the observation time $\tau$ during measurment, such that the time is $t=t_w + \tau$. With the  strain imposed starting at $t=t_w$, 
%we have $\bar{\epsilon}(t)=\Theta(t-t_w)\epsilon(t)$ ($\Theta$ is the Heaviside step function), and 
Eq.(\ref{eq:LVE_G}) becomes 
\begin{equation} 
\begin{split}
\label{eq:}
\sigma(t) = \int_{t_w}^{t} G(t,t'){\epsilon}(t')dt'.  \\ 
\end{split}
\end{equation}
Using the change of variables, $\tau=t-t_w$ and $\tau'=t'-t_w$,
\begin{equation} 
\begin{split}
\label{eq:sigmatw+t}
\sigma(t_w+\tau) = \int_{0}^{\tau} G(t_w+\tau,t_w+\tau'){\epsilon}(t_w+\tau')d\tau'.  \\ 
\end{split}
\end{equation}
One approach to circumvent the complexity of the two time scales is to use observation times $\tau$ much smaller than time scale associated with the change in rheological properties. For such a measurement time,  $G(t_w+\tau,t_w+\tau')\simeq G(t_w,t_w+\tau'-\tau)$ obeys time translational invariance for $\tau$. We denote the resulting dynamic modulus as $G_{t_w}(\tau-\tau')\equiv G(t_w,t_w+\tau'-\tau)$. Then Eq.(\ref{eq:sigmatw+t}) is approximated as,  
\begin{equation} 
\begin{split}
\label{eq:LVE_G2}
\sigma_{t_w}(\tau) \simeq \int_0^{\tau} G_{t_w}(\tau-\tau')\epsilon_{t_w}(\tau')d\tau', \\ 
\end{split}
\end{equation}
where $\sigma_{t_w}(\tau) \equiv \sigma(t_w+\tau)$ and $\epsilon_{t_w}(\tau) \equiv \epsilon(t_w+\tau)$.
Once we approximate the modulus to have time translational invariance for $\tau$, one can obtain the storage and loss modulus for waiting time $t=t_w$ using the same procedure as for the equilibrium case. Repeating this procedure for different $t_w$, we  obtain the $t_w$-dependent material properties.
We remark that the assumption that the observation time $\tau$ is appreciably smaller than the dynamics of the glassy material is not apriori justified and must be checked posteriorly.

An alternative way to obtain the time-dependent material properties during aging, which does not require repeated analysis for different waiting times $t_w$, is to generalize the complex modulus $G(\omega)$ to time-dependent sectra~\cite{fielding2000aging} (Appendix \ref{sec:Hilbert}). The viscoelastic spectra explicitly represent the time-varying material properties, but their computation from experiments is not straightforward. We introduce, in section~\ref{sec:AC}, the instantaneous complex modulus to characterize the rheology of aging materials. We show in Appendix \ref{sec:Hilbert} that the instantaneous complex modulus and the viscoelastic spectra are closely related.
The instantaneous complex modulus does not require the assumption for the observation time-scale and thus captures the full spectrum of the aging material. %The instantaneous complex modulus can be obtained from the active rheology without repeated experiments for different $t_w$.  

\section{\label{sec:Relaxation} Decomposition in dynamic modes.}

We study the relaxation dynamics of Eq.(\ref{eq:ME2}) to the asymptotic solutions for equilibrium and aging regime by defining eigenmodes and eigenvalues. First, we  make the transformation $q_b(E,t)=p_b(E,t) e^{-\beta E/2}/\sqrt{\rho(E)}$, to transform the operator Hermitian, and rewrite Eq.(\ref{eq:ME2}) as
\begin{subequations}
\label{eq:ME2transformed}
  \begin{align} 
  \label{eq:ME2atransformed}
   &\frac{1}{\Gamma_0}\frac{\partial  q_b(E,t)}{\partial t}=-q_b(E,t) e^{-\beta E}+P_u(t)\sqrt{\rho(E)}e^{-\beta E/2} ,\\
   \label{eq:ME2btransformed}
   &\frac{1}{\Gamma_0}\frac{\partial P_u(t)}{\partial t}  =- P_u(t) + \int_0^{\infty}dE q_b(E,t)\sqrt{\rho(E)} e^{-\beta E/2}.
  \end{align}
\end{subequations}
%The eigenvalues do not change under the transformation. 
We introduce eigenfunctions $q_\lambda^{b}(E)$ and $P_\lambda^{u}$
of the linear operator defined in Eq.(\ref{eq:ME2transformed}). 
These eigenfunctions obey
\begin{subequations}
\label{eq:EigenEqs}
  \begin{align} 
  \label{eq:EigenEqsa}
   &-\frac{1}{\Gamma_0}\lambda q_\lambda^{b}(E)=-q_\lambda^{b}(E) e^{-\beta E}+\sqrt{\rho(E)}e^{-\beta E/2} P_\lambda^{u},\\
   \label{eq:EigenEqsb}
   &-\frac{1}{\Gamma_0} \lambda P_\lambda^{u}  =- P_\lambda^{u} + \int_0^{\infty}dE \sqrt{\rho(E)}q_\lambda^{b}(E) e^{-\beta E/2}.
  \end{align}
\end{subequations}
where $\lambda$ denotes the corresponding eigenvalue.

\begin{figure}
\begin{center}
%\scalebox{1.5}[1.5]{
\includegraphics[width=0.4\textwidth]{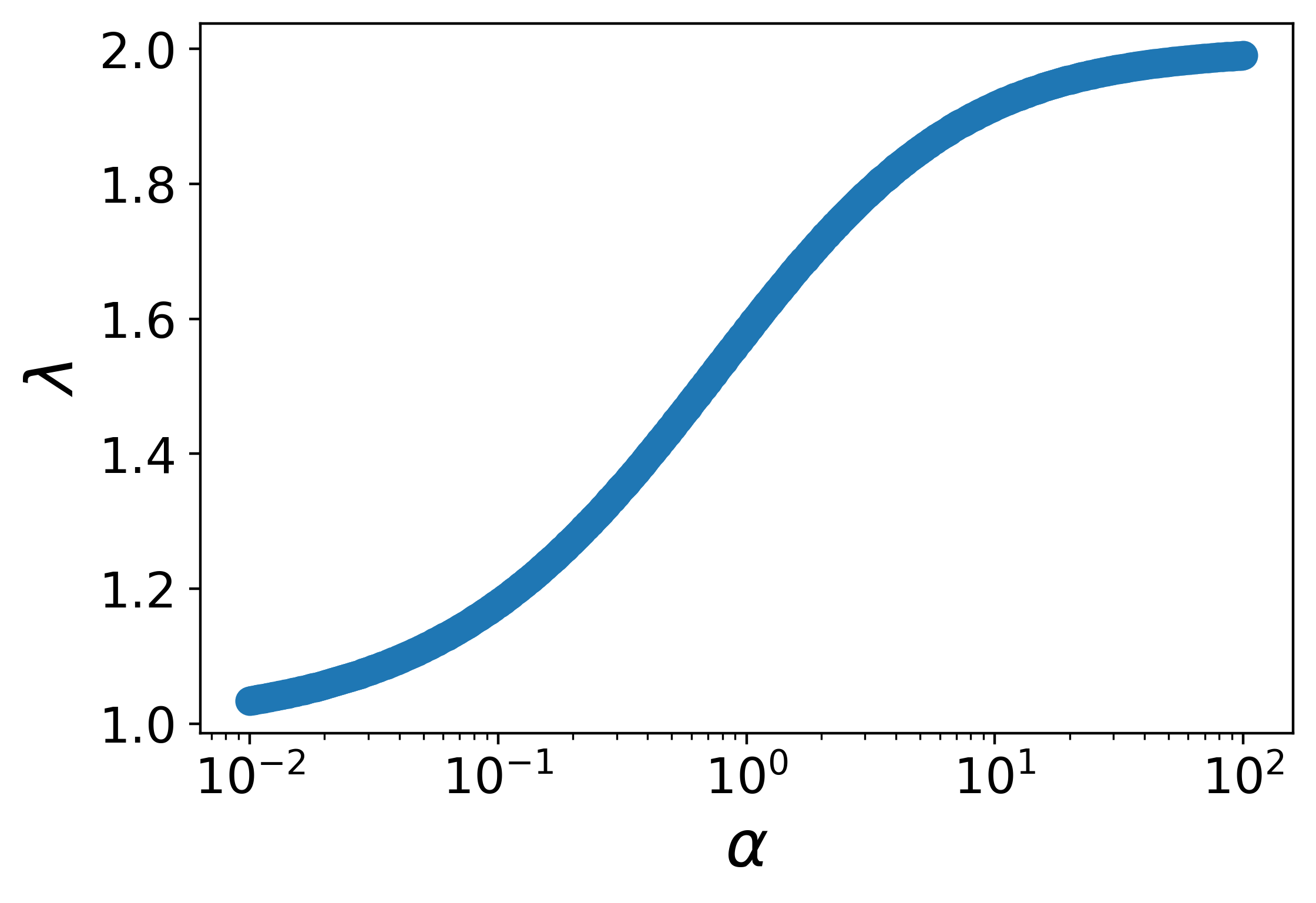}
\end{center}
\caption{\label{Fig:LambdavsAlfa} Eigenvalue $\lambda$ as a function of $\alpha$ obtained by numerically solving Eq.(\ref{eq:isoeigen}). $\Gamma_0$ is set to unity. $\lambda$ determines the relaxation rate to the asymptotic solutions in equilibrium and aging regime (see Fig.\ref{Fig:Puplots}). 
}
\end{figure}

We can eliminate $q_\lambda^b$ from Eq. (\ref{eq:EigenEqs}) which leads to the condition
\begin{equation} 
\begin{split}
\label{eq:}
\Big(1-\frac{1}{1-\lambda/\Gamma_0}\int_0^{\infty}dE \frac{\rho(E)e^{-\beta E}}{e^{-\beta E }-\lambda/\Gamma_0}\Big)P_\lambda^{u} =0.
\end{split}
\end{equation}
In order to find the eigenfunctions, we distinguish two cases. \\
Case (I): $P_\lambda^{u}=0$. In this case 
Eq.(\ref{eq:EigenEqs}) reduces to 
\begin{subequations}
\label{eq:}
  \begin{align} 
  \label{eq:}
   &-\frac{1}{\Gamma_0}\lambda q_\lambda^{b}(E)=-q_\lambda^{b}(E) e^{-\beta E},\\
   \label{eq:EigenEqsb2}
   &0=\int_0^{\infty}dE \sqrt{\rho(E)}q_\lambda^{b}(E) e^{-\beta E/2}.
  \end{align}
\end{subequations}
This can be solved by the ansatz, $q_\lambda^{b}(E)=a \delta(E-E_\lambda) + \delta' (E-E_\lambda)$, where $a$ is a constant. From Eq.(\ref{eq:EigenEqsb2}) we obtain,
\begin{equation} 
\begin{split}
\label{eq:}
a =\frac{\beta-\beta_0}{2},
\end{split}
\end{equation}
leading to 
\begin{equation} 
\begin{split}
\label{eq:}
q_\lambda^{b}(E) = \frac{\beta-\beta_0}{2} \delta(E-E_\lambda) + \delta'(E-E_\lambda),
\end{split}
\end{equation}
with the eigenvalues, $\lambda=\Gamma_0e^{-\beta E}$. \\
Case (II): $P_\lambda^u \neq 0$ and $\int_0^{\infty}dE \frac{\rho(E)e^{-\beta E}}{e^{-\beta E }-\lambda/\Gamma_0}=1-\lambda/\Gamma_0$. 
Using the variable transform $x=e^{-\beta E}$, we find 
\begin{equation} 
\begin{split}
\label{}
&\int_0^{\infty}dE \frac{\rho(E)e^{-\beta E}}{e^{-\beta E }-\lambda/\Gamma_0} = \frac{\beta_0}{\beta}\int_0^1 dx  \frac{x^{\frac{\beta_0}{\beta}}}{x-\lambda/\Gamma_0} \\
%\frac{\beta_0}{\beta} \Big(\frac{\lambda}{\Gamma_0}\Big)^{\frac{\beta_0}{\beta}-1} \int_0^{\Gamma_0/\lambda} dy \frac{y^{\frac{\beta_0}{\beta}}}{y-1}\\
%&=\frac{1}{\lambda}\frac{\beta_0}{\beta}\frac{1}{1+\beta_0/\beta}  
&=-\frac{\beta_0}{\beta}\frac{\Gamma_0}{(1+\beta_0/\beta)\lambda}\mathstrut_2\mathrm{F}_1  \Big(1,\frac{\beta_0}{\beta}+1,\frac{\beta_0}{\beta}+2,\frac{\Gamma_0}{\lambda}\Big) \quad ,
\end{split}
\end{equation}
where $\mathstrut_2\mathrm{F}_1$ is the Hypergeometric function~\cite{abramowitz1964handbook}. 
%In the first change of the variable, we used $x=e^{-\beta E}$.
Therefore the corresponding eigenvalue obeys the equation:
\begin{equation} 
\begin{split}
\label{eq:isoeigen}
\frac{\alpha}{1+\alpha}  \mathstrut_2\mathrm{F}_1  \Big(1,\alpha+1,\alpha+2,\frac{\Gamma_0}{\lambda}\Big) = \frac{\lambda}{\Gamma_0}\Big(\frac{\lambda}{\Gamma_0}-1\Big),
\end{split}
\end{equation}
where $\alpha=\beta_0/\beta$. 
%We show the normalized distributions of the eigenvalues in Fig.\ref{Fig:eigenvalues}. 
Because $P_\lambda^{u}=0$ for case (I), the relaxation dynamics of $P_u(t)$ is fully determined by the eigenvalue satisfying Eq.(\ref{eq:isoeigen}), which depends on $\alpha$. Fig.\ref{Fig:LambdavsAlfa} shows the eigenvalue $\lambda$ as a function of $\alpha$. 

\section{\label{sec:agingsol} Solutions of dynamic equations using  Laplace transforms.}
In this Appendix, we solve Eq.(\ref{eq:ME2}) using the Laplace transform and obtain asymptotic solutions for long time. Because of the conservation of probabilities, $P_u(t) + \int_0^{\infty} dE' p_b(E',t) =1$, Eq.(\ref{eq:ME2}) can be written in one equation,
\begin{equation} 
\begin{split}
\label{eq:LapP(E,t)}
\frac{1}{\Gamma_0}\frac{d}{dt} p_b(E,t)=&-e^{-\beta E}  p_b(E,t) \\
&-\rho(E) \int_{0}^{\infty} p_b(E',t)dE' + \rho(E).
\end{split}
\end{equation}
We take the Laplace transform of Eq.(\ref{eq:LapP(E,t)}) with respect to $t$ and solve for ${p}_b(E,s)$, 
\begin{equation} 
\begin{split}
\label{eq:P(E,s)1}
{p}_b(E,s)=&-\frac{\rho(E)C(s)}{s/\Gamma_0 + e^{-\beta E}} +\frac{ p_b(E,0)/\Gamma_0}{s/\Gamma_0 + e^{-\beta E}} \\
&+\frac{ \rho(E)}{(s/\Gamma_0 + e^{-\beta E})s},
\end{split}
\end{equation}
where 
\begin{equation} 
\begin{split}
\label{eq:P(E,s)2}
C(s)=\frac{\int_{0}^\infty dE'\frac{ p_b(E',0)/\Gamma_0 + }{s/\Gamma_0 + e^{-\beta E'}}+ \int_{0}^\infty dE'\frac{\rho(E')}{(s/\Gamma_0 + e^{-\beta E'})s}}{1+\int_{0}^{\infty}dE'\frac{\rho(E')}{s/\Gamma_0 + e^{-\beta E'}}}.
\end{split}
\end{equation}
Eq.(\ref{eq:P(E,s)1}-\ref{eq:P(E,s)2}) with ${P}_u(s)=1/s-\int_0^{\infty} dE {p}_b(E,s)$ give the complete solution of Eq.(\ref{eq:ME2}) in Laplace space.

%{\it Asymptotic solution for equilibrium and aging regime.} 
We first derive the expression of $P_u(s)$ for $s \rightarrow 0$. 
Integrating Eq.(\ref{eq:P(E,s)1}) for $E$ to obtain,
\begin{equation}
\begin{split}
\label{eq:Pb(s)}
{P}_b(s)&= -C(s) Q_\rho(s) + Q_0(s) +\frac{1}{s}Q_\rho(s), 
\end{split}
\end{equation}
where
\begin{equation}
\begin{split}
\label{eq:}
Q_\rho(s) \equiv \int_0^\infty dE  \frac{\rho(E)}{s/\Gamma_0 + e^{-\beta E}};
\end{split}
\end{equation}
\begin{equation}
\begin{split}
\label{eq:}
Q_0(s) \equiv \int_0^\infty dE  \frac{p_b(E,0)/\Gamma_0}{s/\Gamma_0 + e^{-\beta E}};
\end{split}
\end{equation}
and 
\begin{equation} 
\begin{split}
\label{eq:C(s)}
C(s)=\frac{Q_\rho(s)}{s(1+Q_\rho(s))}+\frac{Q_0(s)}{1+Q_\rho(s)}.
\end{split}
\end{equation}
$P_b(s)$ simplifies to
\begin{equation} 
\begin{split}
\label{}
{P}_b(s) = \frac{Q_\rho(s)}{s(1+Q_\rho(s))}+\frac{Q_0(s)}{1+Q_\rho(s)},
\end{split}
\end{equation}
and 
\begin{equation} 
\begin{split}
\label{eq:Pu0(s)}
{P}_u(s) &= \frac{1}{s}-{P}_b(s)\\
&=\frac{1}{s}\frac{1}{1+Q_\rho(s)} - \frac{Q_0(s)}{1+Q_\rho(s)}.
\end{split}
\end{equation}
The term containing $Q_0(s)$ in the second line of Eq.(\ref{eq:Pu0(s)}) is the contribution from the initial distribution giving subordinate contribution for long time. Here it is set to $0$ because $p_b(E,0)=0$, leading to
%the contribution from the initial distribution, gives subordinate contribution for long time ($s \rightarrow 0$), thus,
\begin{equation} 
\begin{split}
\label{eq:Pu(s)}
{P}_u(s) = \frac{1}{s}\frac{1}{1+Q_\rho(s)}.
\end{split}
\end{equation}
One can explicitly evaluate $Q_\rho(s)$ for $s \rightarrow 0$ as follows for equilibrium case (I) and aging case (II).

Equilibrium case (I). For the equilibrium case one can expand $Q_{\rho}(s)$ as follows for $s \rightarrow 0$,
\begin{equation} 
\begin{split}
\label{}
Q_\rho(s)&=\int_0^\infty dE  \frac{\beta_0 e^{-\beta_0 E}}{s/\Gamma_0 + e^{-\beta E}} \\
&\simeq \frac{\beta_0}{\beta_0-\beta}-\frac{s}{\Gamma_0}\frac{\beta_0}{\beta_0-2\beta} + O(s^2).
\end{split}
\end{equation}
We substitute the first term of the expansion into Eq.(\ref{eq:Pu(s)}) to obtain,
\begin{equation} 
\begin{split}
\label{}
P_u(s) \simeq \frac{\beta_0/\beta-1}{s(2\beta_0/\beta-1)}.
\end{split}
\end{equation}
Inverting to the real space, we have,
\begin{equation} 
\begin{split}
\label{}
P^{eq}_u =  \frac{\alpha-1}{2\alpha-1},
\end{split}
\end{equation}
where $\alpha=\beta_0/\beta >1$.

Aging case (II). For aging case, we first make variable transforms to extract the power law form of $s$: 
\begin{equation} 
\begin{split}
\label{}
Q_\rho(s)&=\int_0^\infty dE  \frac{\beta_0 e^{-\beta_0 E}}{s/\Gamma_0 + e^{-\beta E}} \\
&= \frac{\beta_0}{\beta}\int_0^1 dx  \frac{x^{\frac{\beta_0}{\beta}-1}}{s/\Gamma_0 +x} \\
&= \frac{\beta_0}{\beta} \Big(\frac{s}{\Gamma_0}\Big)^{\frac{\beta_0}{\beta}-1}\int_0^{\Gamma_0/s} dy  \frac{y^{\frac{\beta_0}{\beta}-1}}{1+y}.
\end{split}
\end{equation}
In the second line, we used the change of variables $x=e^{-\beta E}$ and the third line, $y=x \Gamma_0/s$. 
In the limit of $s\rightarrow 0$, we can extend the upper bound of the integral in the third line to $\infty$:
\begin{equation} 
\begin{split}
\label{}
\int_0^{\infty} dy  \frac{y^{\frac{\beta_0}{\beta}-1}}{1+y} = \pi \csc \Big(\frac{\beta_0}{\beta}\pi \Big).
\end{split}
\end{equation}
Thus, in the limit of $s \rightarrow 0$,
\begin{equation} 
\begin{split}
\label{}
Q_\rho(s) \simeq \frac{\beta_0}{\beta} \Big(\frac{s}{\Gamma_0}\Big)^{\frac{\beta_0}{\beta}-1}\pi \csc \Big(\frac{\beta_0}{\beta}\pi \Big).
\end{split}
\end{equation}
Noting that $\beta_0/\beta-1<0$ in the aging regime, $1+Q_\rho(s)\simeq Q_\rho(s) $ for  $s \rightarrow 0$.
From Eq.(\ref{eq:Pu(s)}),
\begin{equation} 
\begin{split}
\label{}
{P}_u(s) \simeq   \frac{1}{s Q_\rho(s)} = \frac{\beta \Gamma_0 \sin\big(\frac{\beta_0}{\beta}\pi \big)}{\beta_0(s/\Gamma_0)^{\frac{\beta_0}{\beta}} \pi  }.
\end{split}
\end{equation}
By taking the inverse Laplace transform we obtain the result for long time, 
\begin{equation} 
\begin{split}
\label{eq:PuagingApp}
P_u(t) =  \frac{\sin \big(\alpha\pi \big)}{\alpha \pi \Gamma \big[\alpha\big] } \big(\Gamma_0 t\big)^{\alpha-1},
\end{split}
\end{equation}
where $\alpha=\beta_0/\beta <1$.

One can find complete solutions for special cases, infinite temperature ($\beta = 0$) and zero temperature ($\beta = \infty$).
For the infinite temperature case, solving Eq.(\ref{eq:P(E,s)1}-\ref{eq:P(E,s)2}) and taking the inverse Laplace transform, we obtain,
% \begin{equation} 
% \begin{split}
% \label{}
% p(E,t)=\frac{1}{2}\Big(\rho(E) + e^{-2\Gamma t}(-1+2 P(0))\rho(E) + 2 e^{-\Gamma t}\big(p(E,0)- P(0)\rho(E)\big) \Big).
% \end{split}
% \end{equation}
\begin{equation} 
\begin{split}
\label{}
P_b(t)&=\frac{1}{2} \big(1+ e^{-2\Gamma_0 t} (-1+2 P_b(0)) \big);\\
P_u(t)&=\frac{1}{2} \big(1- e^{-2\Gamma_0 t} (-1+2 P_b(0)) \big).
\end{split}
\end{equation}
For the zero temperature case,
solving Eq.(\ref{eq:P(E,s)1}-\ref{eq:P(E,s)2}) and taking inverse Laplace transform, we obtain,
\begin{equation} 
\begin{split}
\label{}
P_b(t)&=1-P_u(0) e^{-\Gamma_0 t};\\
P_u(t)&=P_u(0) e^{-\Gamma_0 t}.
\end{split}
\end{equation}
This suggests that the dynamics is completely frozen for zero temperature. 

\section{\label{sec:Change_of_Elastic} Change of elasticity in aging regime}
\begin{figure}
\begin{center}
%\scalebox{1.5}[1.5]{
\includegraphics[width=0.5\textwidth]{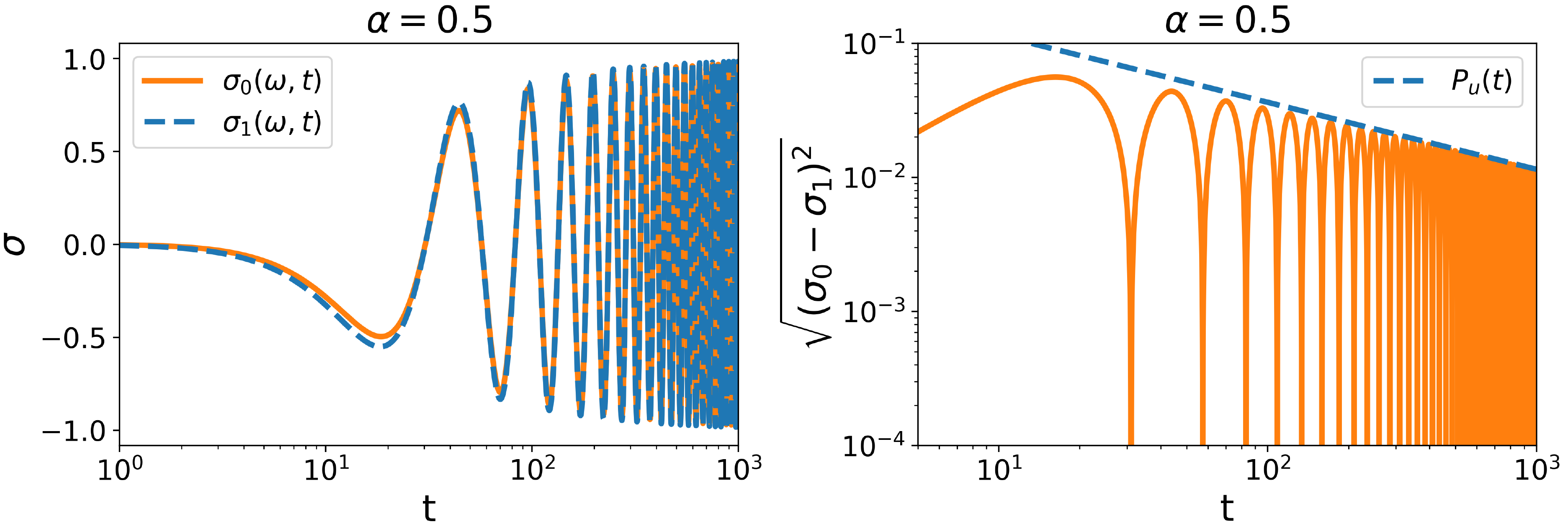}
\end{center}
\caption{\label{Fig:comp}  {\bf Left:} Comparison of the resulting stresses $\sigma_1$ and $\sigma_0$ obtained from the model with effective elastic modulus $G_c= G_0/(1 - P_u(t))$ (blue dashed line) and $G_c= G_0$ (orange solid line), respectively. The value of frequency $\omega=\pi/5$ used in this example is the highest frequency presented in Fig.3, for which the difference between $\sigma_1$ and $\sigma_0$ is most pronounced. {\bf Right:} The square root of difference between $\sigma_1$ and $\sigma_0$. The dashed line is $P_u(t)$ determining the decay of the difference. 
%The parameters are identical to Fig.\ref{Fig:AR}a except the initial condition for $P_u(t)$ because of the singularity at $1$ in Eq.(\ref{eq:CE_G}), here the initial value is set to $0.999$. 
}
\end{figure}

% In this Appendix, we show that the neglecting the change in shear elastic modulus in the constitutive equation Eq.(\ref{eq:CE}), $G_c= G_0/(1 - P_u(t))$ has a minor effect on the aging dynamics of protein condensate, justifying our choice to approximate $G_c \simeq G_0$ for the analytical calculation in the main text. The constitutive equation reads:
% \begin{equation} 
% \label{eq:CE_G}
% \dot{\epsilon}(t) = \frac{\sigma(t)}{\eta_0}P_u(t) + \frac{\dot{\sigma}(t)}{G_0}\big(1-P_u(t)\big).
% \end{equation}
%and the corresponding relaxation function is therefore:
%\begin{align} 
%\label{eq:}
%K(t,t') =& G_0\big(1-P_u(t)\big) \\
%&\exp \Big(-\frac{G_0}{\eta_0}\int_{t'}^{t} dt'' %P_u(t'')\big( 1-P_u(t'')\big)\Big ).
%\end{align}
%}
%
%\textcolor{blue}{
In the aging regime  the fraction of bound cross-linker $(1-P_u(t))$ quickly converges towards 1. This can be substantiated by the numerical values of $P_u(t)$ presented in Fig.\ref{Fig:Puplots}b, which are several orders of magnitude smaller than 1. 
In the equilibrium regime, the $P_u(t)$ value is constant, and so any change to $G_0$ would also be constant. As such, the effect of $(1-P_u(t))$ on the modulus $G_0$ does not alter the overall behaviour of the system.
We numerically test the effect of the correction term $P_u(t)$ by imposing a periodic shear strain in the model with $G_c=G_0/(1-P_u(t))$ and $G_c=G_0$, and calculating the resulting stresses, see Fig.\ref{Fig:comp}. We find that the magnitude of difference between the two stresses is bounded by $P_u(t) \to 0$. 

\section{\label{sec:Hilbert} Hilbert transform, analytic signal, and rheology.}
We refer Ref.~\cite{king2009hilbert_1,king2009hilbert} for the theory and various applications with a comprehensive table of Hilbert transform. We discuss here the basic definition of Hilbert transform and analytic signal, and the connection to rheology. 
The Hilbert transform of a function, $f(t)$, is defined as 
\begin{equation} 
\begin{split}
\label{}
\mathcal{H}[f](t) = \frac{1}{\pi}p.v.\int_{-\infty}^{\infty}\frac{f(t')}{t-t'}dt', 
\end{split}
\end{equation}
where $p.v.$ denotes Cauchy  principle value. 
Fourier transform ($\mathcal{F}$) of Hilbert transformed signal is the $\pm 90$ degrees phase shift, depending on the sign of the frequency $\omega$, of the original signal, namely,
\begin{equation} 
\begin{split}
\label{eq:HilbertFourier}
\mathcal{F}\big[\mathcal{H}[f]\big](\omega) = -i \text{sgn}(\omega) \mathcal{F}[f](\omega),
\end{split}
\end{equation}
where sgn is signum function. 
Using the Hilbert transform, analytic representation of $f(t)$ is
\begin{equation} 
\begin{split}
\label{eq:defAS}
f_a(t)=f(t) +i \mathcal{H}[f](t).
\end{split}
\end{equation}
In the context of the active rheology of aging material, the following theorem is useful. \\

{\it Bedrosian's theorem}~\cite{bedrosian1962product}: Suppose a low-pass signal, $l(t)$, and high-pass signal, $h(t)$, have Fourier transforms $L(\omega)$ and $H(\omega)$, respectively, where $L(\omega)=0$ for $|\omega|>\omega_0$ and $H(\omega)=0$ for $|\omega|<\omega_0$. Then,
\begin{equation} 
%\begin{split}
\label{}
\mathcal{H}[l(t)h(t)] = l(t)\mathcal{H}[h(t)].
\end{equation}
Namely, the product of a low-pass and a high-pass signal with non-overlapping spectra is obtained by the product of the low-pass signal and the Hilbert transform of the high-pass signal. In the context of rheology, Bedrosian's theorem requires the spectra of the aging to have a maximum spectrum smaller than the frequency of input sinusoidal shear strain.

{\it Time-varying viscoelastic spectrum.}
We illustrate the connection of the analytic signal to the time-dependent rheology of aging materials. 
Let us consider the relation between the stress and strain-rate of a material with a relaxation function $K(t,t')$:
\begin{equation} 
\begin{split}
\label{eq:responce}
\sigma(t) = \int_{0}^t K(t,t') \dot{{\bar{\epsilon}}}(t') dt'.  
\end{split}
\end{equation}
We apply the sinusoidal strain having frequency $\omega$ starting at $t=t_w$: $\bar{\epsilon}(\omega,t,t_w) = \Theta(t-t_w)\epsilon(t)$ where $\epsilon(t)=\Re[\epsilon_0 e^{i (\omega t + \varphi_0)}]$ and $\Theta(t)$ is Heaviside step function. Substituting ${\epsilon}(\omega,t,t_w)$ to Eq.(\ref{eq:responce}) leads to  
\begin{equation} 
\begin{split}
\label{eq:sigmaG*}
\sigma(\omega,t, t_w) =\Re [\epsilon_0 e^{i(\varphi_0 + \omega t)}{G}^*(\omega,t,t_w)],
\end{split}
\end{equation}
where 
\begin{equation} 
\begin{split}
\label{eq:G*}
{G}^*(\omega,t,t_w)\equiv &i \omega \int_{t_w}^{t} e^{-i\omega (t-t')}K(t,t')dt'\\
&+ e^{-i \omega (t-t_w)}K(t,t_w).
\end{split}
\end{equation}
${G}^*(\omega,t,t_w)$ is the time-varying viscoelastic spectrum~\cite{fielding2000aging}. 

We show that the time-varying viscoelastic spectrum may be obtained from the method of analytic signal. The analytic signal of the input strain, $\Re[\epsilon_0 e^{i (\omega t + \varphi_0)}]$, is $\epsilon_a(t) = \epsilon_0 e^{i (\omega t + \varphi_0)}$. Taking the Hilbert transform of Eq.(\ref{eq:sigmaG*}),
\begin{equation} 
\begin{split}
\label{}
\mathcal{H}[\sigma(\omega,t,t_w)] = \Re \Big[\mathcal{H} \big[\epsilon_a(\omega,t) {G}^*(\omega,t,t_w)\big] \Big].
\end{split}
\end{equation}
Assuming the spectra of ${G}^*(\omega,t,t_w)$ for $t$ and spectra of the input shear strain, $\omega$, satisfy the Bedrosian's theorem,
\begin{equation} 
\begin{split}
\label{eq:Himag}
\mathcal{H}[\sigma(\omega,t,t_w)] &= \Re \Big[ \mathcal{H}\big[\epsilon_a(\omega,t) \big] {G}^*(\omega,t,t_w)\Big]\\
 &= \Re \big[ -i \epsilon_a(\omega,t) {G}^*(\omega,t,t_w)\big]\\
 &= \Im \big[ \epsilon_a(\omega,t) {G}^*(\omega,t,t_w)\big].
\end{split}
\end{equation}
Thus, from the definition of analytic signal [Eq.(\ref{eq:defAS})] with Eq.(\ref{eq:sigmaG*}) and Eq.(\ref{eq:Himag}), the analytic signal of $\sigma(\omega,t,t_w)$ is written as 
\begin{equation} 
\begin{split}
\label{}
\sigma_a(\omega,t,t_w) = \epsilon_a(\omega,t) {G}^*(\omega,t,t_w).
\end{split}
\end{equation}
Therefore the definition of the instantaneous complex modulus, Eq.(\ref{eq:Gtdef}), gives: 
\begin{equation} 
\begin{split}
\label{}
G(\omega,t,t_w) \equiv \frac{\sigma_a(\omega,t,t_w)}{\epsilon_a(\omega,t)} = {G}^*(\omega,t,t_w). 
\end{split}
\end{equation}
This shows that, under the Bedrosian's theorem, the instantaneous complex modulus and the viscoelastic spectra are identical.

It may be instructive to consider the simple Maxwell fluid. Because the Hilbert transform is a linear transform, we can write the constitutive equation of simple Maxwell fluid using analytic signal,
\begin{equation} 
\begin{split}
\label{eq:}
\dot{\epsilon}_a(t) = \frac{\sigma_a(t)}{\eta_0} + \frac{\dot{\sigma}_a(t)}{G_0}. 
\end{split}
\end{equation}
Let us consider the input stress $\sigma(\omega,t)=\sigma_0 \cos(\omega t)$. The analytic signal of $\sigma(\omega,t)$ is $\sigma_a(\omega,t)=\sigma_0e^{i \omega t}$. The explicit integration of right-hand side, setting integration constant $0$, to obtain ${\epsilon}_a(\omega,t)$ leads to ${\epsilon}_a(\omega,t)=\sigma_0e^{i\omega t}(1/G_0-i/\big(\eta_0 \omega)\big)$.  Therefore $G(\omega,t) = \sigma_a(\omega,t)/{\epsilon}_a(\omega,t) = 1/\big(1/G_0-i/(\eta_0 \omega)\big)$, which is the complex modulus of Maxwell fluid which does no have time dependence. Therefore, Eq.(\ref{eq:Gtdef}) recovers the definition of conventional complex modulus.

\section{\label{sec:FRR} Aging fluctuation-dissipation theorem for Maxwell glass}

We first obtain the strain-stress response function, $\chi(t,t')$, for the constitutive equation, Eq.(\ref{eq:CE}). 
%Noting that $\bar{\epsilon}(t)$ and $\sigma(t)$ are relative to the time for the application of stress, $t=0$, 
\begin{equation} 
\begin{split}
\label{eq:}
{\epsilon}(t) &= \int_0^{t}
\Theta(t-t')\Big(\frac{P_u(t')}{\eta_0} + \frac{1}{G_0}\frac{d}{dt'}\Big)\sigma(t')dt' \\ 
&=  \int_0^{t} \Big( \Theta(t-t')\frac{P_u(t')}{\eta_0} + \frac{2\delta(t-t')}{G_0}\Big) \sigma(t')dt',
\end{split}
\end{equation}
where $\Theta(t)$ is the Heaviside step function. The factor $2$ in front of the delta function is to account for the boundary.
Therefore the response function is given, as
\begin{equation} 
\begin{split}
\label{eq:defchi}
\chi(t,t') = \Theta(t-t') \frac{P_u(t')}{\eta_0} + \frac{2\delta(t-t')}{G_0}.
\end{split}
\end{equation}

On the other hand, using Eq.(\ref{eq:D(t)}) and the constant $4k_BT/G_0$, we compute
\begin{equation} 
\begin{split}
\label{eq:}
&\Theta(t-t')\frac{d}{dt'}\langle \Delta x^2(t') \rangle \\
&= \Theta(t-t')\Big(2D_0 P_u(t')+  \frac{d}{dt'} \frac{4k_BT}{G_0}\Big)\\
&= 2k_BT \Big(\Theta(t-t')\frac{P_u(t')}{\eta_0} + \frac{2\delta(t-t')}{G_0} \Big),
\end{split}
\end{equation}
where we used integration by parts from the second line to the third line and the Einstein relation $D_0\eta_0=k_BT$~\cite{zwanzig2001nonequilibrium}. 
Therefore we obtain the fluctuation-response relation, Eq.(\ref{eq:GFDT}).

The response function $\chi(t,t')$ is related to the dynamic modulus $G(t,t')$ by inverse, thus uniquely determined. To see this we notice that the shear strain $\epsilon(t)$ is written using Eq.(\ref{eq:LVE_G}) and Eq.(\ref{eq:LVE_chi}) as

\begin{equation} 
\begin{split}
\label{eq:ep-ep}
\epsilon(t) &= \int_0^t dt' \chi(t,t') \int_0^{t'} dt'' G(t',t'')\epsilon(t'')\\
&=\int_0^t dt'' \epsilon(t'') \int_{t''}^t dt' \chi(t,t')G(t',t'').
\end{split}
\end{equation}
By direct calculation using Eq.(\ref{eq:defchi}) and Eq.(\ref{eq:K}) and using that the general form of $K(t,t')$ in our model [Eq.(\ref{eq:Ktt'nu})] has exponential form, we obtain
\begin{equation} 
\begin{split}
\label{eq:pseudo1}
\int_{t''}^{t}dt'\chi(t,t')G(t',t'') &=  2\delta(t-t''),
\end{split}
\end{equation}
leading to the consistent expression for Eq.(\ref{eq:ep-ep}). Note that the factor $2$ accounts for the integration of the delta function at the boundary. Eq.(\ref{eq:pseudo1}) shows that $\chi(t,t')$ and $G(t,t')$ are related by inverse and uniquely determined. 
% \begin{equation} 
% \begin{split}
% \label{eq:}
% \int_{t''}^{t}dt'G(t,t')\chi(t',t'') &=  2\delta(t-t'').
% \end{split}
% \end{equation}
% Consequently, we have
% \begin{equation} 
% \begin{split}
% \label{eq:}
% \int_{t'''}^{t}dt'' \int_{t''}^{t}dt'\chi(t,t')G(t',t'') \chi(t'',t''')=\chi(t,t''')
% \end{split}
% \end{equation}
% and 
% \begin{equation} 
% \begin{split}
% \label{eq:pseudo4}
% \int_{t'''}^{t}dt'' \int_{t''}^{t}dt'G(t,t')\chi(t',t'') G(t'',t''')=G(t,t''').
% \end{split}
% \end{equation}
%Eq.(\ref{eq:pseudo1})-Eq.(\ref{eq:pseudo4}) shows that $\chi(t,t')$ and $G(t,t')$ are related by pseudo-inverse and uniquely determined. 

\section{\label{sec:NumEq1} Numerical procedure to solve the trap model for the protein condensates}
In order to solve Eq.(\ref{eq:ME2}) numerically, we first rewrite Eq.(\ref{eq:ME2}) as Eq.(\ref{eq:LapP(E,t)}) using the conservation of the probabilities for $p_b(E,t)$ and $P_u(t)$. The unit of time is $1/\Gamma_0$ and we set $\Gamma_0=1$. We discretize the time and energy using sufficiently small steps, here we use the time step $\Delta t=0.01$ and the step for the energy $\Delta E=0.02\ (k_BT)$. For the numerical computation it is necessary to introduce the cut-off for the energy. We set the maximum energy to be $100\ (k_BT)$ in the numerical computation. The integral for the energy is simply the sum of the probability density, multiplied by $\Delta E$, in the discretized computation.  We use the Euler method for the integration over time.

\section{\label{sec:NAR} Numerical procedures to compute instantaneous complex modulus. }
\begin{figure}[t]
\begin{center}
%\scalebox{1.5}[1.5]{
\includegraphics[width=0.45\textwidth]{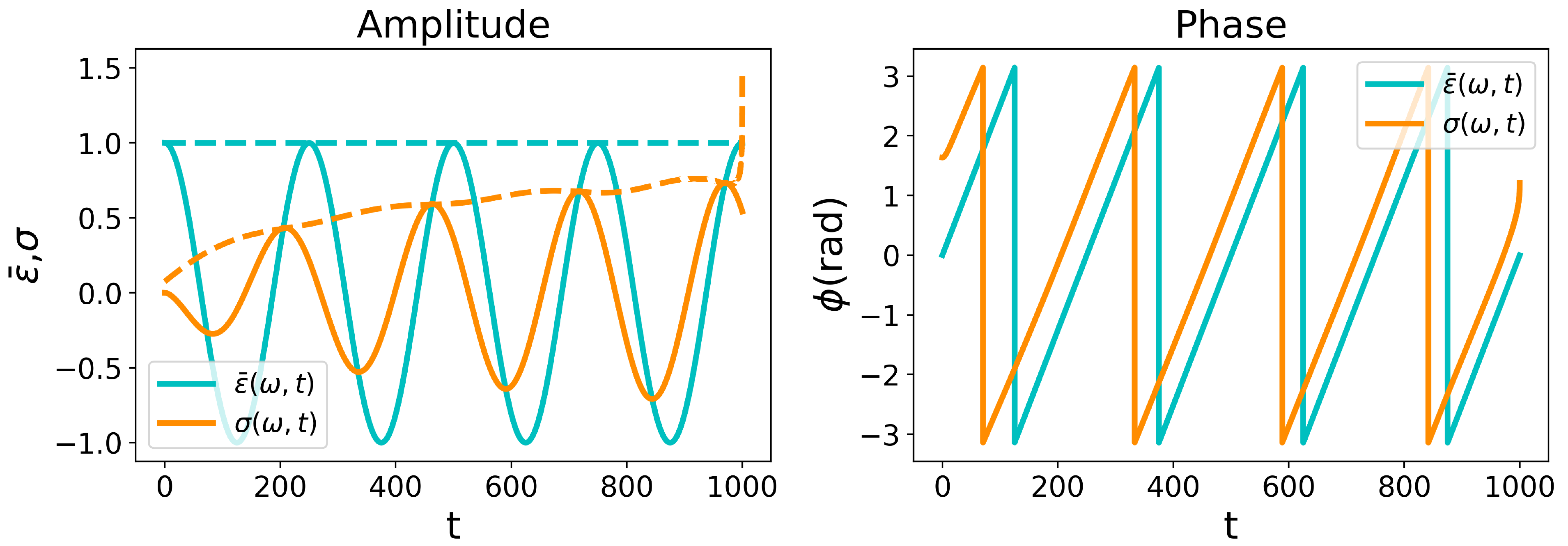}
%}
\end{center}
\caption{\label{Fig:NumericsHilbert} An example of the instantaneous amplitude and phase extraction using analytic signal. {\bf Left:} Amplitude extraction from the data using analytic signal. The solid line in cyan is the input strain $\bar{\epsilon}(\omega,t)$, and the solid orange line is the output stress $\sigma(\omega,t)$. Dashed curves are the instantaneous amplitude of $\bar{\epsilon}(\omega,t)$ and $\sigma(\omega,t)$, computed using analytic signal. {\bf Right:} Instantaneous phase for $\epsilon(\omega,t)$ (cyan) and $\sigma(\omega,t)$ (orange), computed using analytic signal. }
\end{figure}

% In order to compute the instantaneous complex modulus $G(\omega,t,t_w)$, we obtain the instantaneous amplitude and phase of the input and output signals using analytic signal approach. We used the python package "scipy.signal.hilbert"~\cite{2020SciPy-NMeth} to extract the instantaneous amplitude and phase for the input shear strain and output shear stress (Fig.\ref{Fig:NumericsHilbert}). 

To compute the instantaneous complex modulus $G(\omega,t,t_w)$, we employed the analytic signal approach to obtain the instantaneous amplitude and phase of the input and output signals. Specifically, we utilized the Python package "scipy.signal.hilbert"~\cite{2020SciPy-NMeth} to extract the instantaneous amplitude and phase for the input shear strain and output shear stress, as illustrated in Fig.\ref{Fig:NumericsHilbert}. This method allowed us to accurately capture the time-varying behavior of the signals and determine the complex modulus at any given time and frequency.

The computation of the instantaneous complex modulus $G(\omega,t,t_w)$, as defined in Eq.(\ref{eq:Gtdef}), requires the input shear strain $\epsilon(\omega,t)$ to span from $t=-\infty$ to $t=\infty$.  Practically, when implementing the numerical computation of instantaneous complex modulus, we extrapolate the input shear strain used in the rheology experiment. Here we extended the imposed sinusoidal shear strain starting $t=t_w$ and ending $t=t_f$: $\epsilon(\omega,t)\Theta(t-t_w)\Theta(t_f-t)$, to the signal from $t=t_w-\tau$ to $t=t_f+\tau$, where $\tau=t_f-t_w$ is the duration of the shear strain. For the output shear stress, we inserted $0$ from $t=t_w-\tau$ to $t=t_w$ and from $t=t_f$ to $t=t_f + \tau$, to adjust the length of the input and output signals. After the extension of the input shear strain and output shear stress we computed Hilbert transform and then extracted back the original, experimentally relevant, part of the signal defined from $t=t_w$ to $t=t_f$. We computed the instantaneous complex modulus using the obtained analytic signal for the input shear strain and output shear stress.

The Hilbert transform, computed using Fourier transform as shown in Eq.(\ref{eq:HilbertFourier}), may produce unwanted oscillations, known as the Gibbs phenomenon, due to the finite discontinuous signal (as illustrated in Fig.\ref{Fig:NumericsHilbert}). To obtain accurate results, we truncated the two edges of the complex modulus, i.e., the initial and final times where the artifact is most prominent. Additionally, we convolved the resulting $G(\omega,t,t_w)$ with a box-kernel whose length was identical to the wavelength of the input shear strain to mitigate the oscillations caused by the Gibbs phenomenon. This step helped to improve the accuracy of our results, shown in Fig.\ref{Fig:AR}.

%\newpage

\bibliography{mybib.bib}
\end{document}